\documentclass[journal,10pt,twocolumn]{IEEEtran} 
\usepackage{cite}
\usepackage{amsmath,amssymb,amsfonts}
\usepackage{algorithmic}
\usepackage{graphicx}
\usepackage{textcomp}
\usepackage{xcolor}
\usepackage{centernot}
\usepackage{epsfig}
\newtheorem{theorem}{Theorem}
\newtheorem{proposition}{Proposition}

\usepackage{graphicx}

\begin{document}

\title{Sparse Activity Detection in Multi-Cell Massive MIMO Exploiting Channel Large-Scale Fading}

\author{Zhilin~Chen,~\IEEEmembership{Member,~IEEE},
Foad~Sohrabi,~\IEEEmembership{Member,~IEEE},
Wei~Yu,~\IEEEmembership{Fellow,~IEEE}
\thanks{This work is supported by the Natural Sciences and Engineering Research Council (NSERC) of Canada.
The authors are with The Edward S. Rogers Sr. Department of Electrical and Computer Engineering, University of Toronto, Toronto, ON M5S 3G4, Canada (e-mails:\{zchen, fsohrabi, weiyu\}@comm.utoronto.ca).
}
}

\maketitle

\begin{abstract}
This paper studies the device activity detection problem in a multi-cell
massive multiple-input multiple-output (MIMO) system, in which the active devices
transmit signature sequences to multiple base stations (BSs) that are connected to a
central unit (CU), and the BSs cooperate across multiple cells to detect the
active devices based on the sample covariance matrices at the BSs. This paper
demonstrates the importance of exploiting the knowledge of channel large-scale
fadings in this cooperative detection setting through a phase transition
analysis, which characterizes the length of signature sequences needed for
successful device activity detection in the massive MIMO regime. It is shown
that when the large-scale fadings are known, the phase transition for the
multi-cell scenario is approximately the same as that of a single-cell system.
In this case, the length of the signature sequences required for reliable activity
detection in the multi-cell system can be made to be independent of the number
of cells through cooperation, in contrast to the case where the large-scale
fadings are not known. Further, this paper considers the case in which the fronthaul links between the
BSs and the CU have capacity constraints and proposes a novel cooperation scheme
based on the quantization of preliminary detection results at the BSs and the
reconstruction of the sample covariance matrices at the CU. Simulations show that the proposed method significantly outperforms the scheme of directly quantizing the sample covariance matrices.
\end{abstract}


\section{Introduction}

The Fifth-Generation (5G) and beyond cellular networks are expected to provide wireless connectivity for massive machine-type communications (mMTC) in which a large number of devices are connected to the network, but the device traffics are sporadic \cite{Bockelmann2016,Dawy2017,Chen2021JSAC}.
A crucial task in mMTC is \emph{sparse activity detection}, which aims to identify the set of active devices in the random access phase based on the signature sequences transmitted in the pilot phase \cite{Liu2018b}. Due to the massive number of devices but limited coherence time, the signature sequences assigned to the devices have to be non-orthogonal, in contrast to the orthogonal sequences used for the random access in the conventional cellular systems. 
The non-orthogonality of the signature sequences in mMTC complicates the task of device activity detection at the base stations (BSs) considerably, because it causes both intra-cell and inter-cell  interference. Although massive multiple-input multiple-output (MIMO) can be used to enhance the sparse activity detection performance of single-cell systems by exploiting spatial dimensions \cite{Liu2018}, the inter-cell interference remains a serious impairment.
This paper explores the use of a cloud radio-access network (C-RAN) architecture to mitigate inter-cell interference. In a C-RAN, the BSs are connected to a central unit (CU) via fronthaul links and can cooperate to perform inter-cell interference cancellation. The main finding of this paper is that if the large-scale fadings from all the devices to all the BSs are known, then by exploiting such knowledge, the performance of a cooperative multi-cell system can have approximately the same scaling in the length of signature sequences and the number of (active) devices as a single-cell system in the massive MIMO regime. Further, practical quantization schemes can be designed to take into account the capacity constraints of the fronthaul links between the BSs and the CU.

The device activity detection problem for mMTC can be formulated as a compressed sensing problem because of the sporadic nature of the device activities. Specifically, device activity can be obtained through joint activity detection and channel estimation using sparse recovery methods. This compressed sensing approach is taken in \cite{Chen2018,Liu2018}, where an approximate message passing (AMP) algorithm is used to recover the device activity and the instantaneous channel state information (CSI) simultaneously. When the BS is equipped with massive MIMO, it is possible to take advantage of the channel hardening effect and to use an alternative approach of jointly estimating the device activity and only the large-scale fading component of the channel of each device (instead of the instantaneous channel realization). This approach, pioneered in \cite{Haghighatshoar2018}, is termed the covariance approach, because it is based on a sample covariance matrix of the received signal. As compared to the compressed sensing approach, the covariance approach has the key advantage that it is capable of detecting a much larger number of active devices at a given pilot length. Specifically, an analytic scaling law derived in \cite{Fengler2019a} shows that the required signature length $L$ for reliably identifying $K$ active devices from $N$ potential devices scales as $L^2=O(K\log^2(N/K))$. The phase transition for the covariance approach is also studied in \cite{ChenZ2020} but from a different perspective, where a numerical optimization problem is used to characterize the minimum value of $L$ needed for reliable activity detection at given $N$ and $K$.

The analyses in \cite{Fengler2019a,ChenZ2020} are obtained assuming a single-cell system without considering inter-cell interference. For multi-cell systems, treating inter-cell interference as noise at full frequency reuse would lead to poor performance. One way to combat the inter-cell interference problem is through multi-cell cooperation. For a multi-cell system consisting of $B$ cells and $N$ potential devices per cell, if all $B$ BSs are connected to a CU via fronthaul links, then they can be considered as a super BS, and accordingly, the multi-cell system becomes a large virtual single-cell system with $BN$ potential users and $BK$ active users.
Based on the scaling law analysis in \cite{Fengler2019a,ChenZ2020}, one may conclude that the signature sequence length $L$ required for successful detection should scale as $L^2=O(BK\log^2(N/K))$.
But this scaling law depends on $B$. For practical networks where $B$ is large, this is undesirable.

The key observation of this paper is the following. The scaling law derived in
\cite{Fengler2019a,ChenZ2020} assumes that the BS does not know the large-scale
fadings, so it needs to jointly detect the device activities and the
large-scale fadings. But in many practical deployment scenarios, the devices
are stationary, so their large-scale fadings are fixed and can be obtained in
advance. This paper shows that in these scenarios, by exploiting the knowledge
of the large-scale fading from the devices to all the BSs, it is possible to
derive an algorithm for device activity detection in the multi-cell setting that
has approximately the same scaling law as a single-cell system, i.e., with a
required signature sequence length for reliable detection that does not depend
on $B$. This is achieved using cooperative detection at the CU that identifies
the active devices in all cells simultaneously. It shows that the knowledge of
large-scale fadings can bring significant benefits to activity detection in the
multi-cell setting.

This paper also studies the practical scenario in which the fronthaul links
between the BSs and the CU have finite capacities. We propose novel
quantization schemes for the multi-cell covariance approach that significantly
reduces the capacity requirements of the fronthaul links as compared to the
naive approach of direct quantization of the covariance matrices.

\subsection{Related Work}
The sparse activity detection with non-orthogonal signature sequences is closely related to a compressed sensing problem. Due to the large-scale nature of this problem in the context of mMTC, the computationally efficient AMP algorithm \cite{Donoho2009} is well suited for solving the compressed sensing problem, and various AMP based device activity detection algorithms have been previously proposed. For single-cell systems, \cite{Liu2018,Chen2018,Sun2019} consider the AMP based joint activity detection and channel estimation, and \cite{Senel2018,Jiang2020} further take the data detection into account. The use of AMP for multi-cell systems has also been studied in \cite{Simeone2016,Chen2019c,Ke2020} by assuming different network architectures. More specifically, \cite{Simeone2016} considers the C-RAN system with capacity-limited fronthaul links, and compares the centralized
and distributed strategies for activity detection under various fronthaul constraints. The work \cite{Chen2019c} studies two network architectures, cooperative MIMO and massive MIMO, and compares their effectiveness in combating the inter-cell interference for activity detection. The cell-free architecture is considered in \cite{Ke2020}, where the AMP based activity detection algorithms are designed for cloud computing and edge computing paradigms. Besides AMP, other compressed sensing based activity detection methods, including Bayesian sparse recovery \cite{Lau2015, Ahn2019}, and dimension reduction based optimization \cite{Shao2020}, have also been discussed in the literature.

The covariance based activity detection has recently attracted extensive research interests due to its capability of detecting many more active devices in a massive MIMO system as compared to the AMP based methods. The covariance approach is originally proposed in \cite{Haghighatshoar2018}, where the activity detection problem is formulated as either a maximum likelihood estimation (MLE) problem, or a non-negative least squares (NNLS) problem, both relying on the sample covariance matrix.
The covariance based method has also been used for
joint activity and data detection \cite{Chen2019a}. The coordinate descent algorithm is used in \cite{Haghighatshoar2018,Chen2019a} for solving the MLE or NNLS problem. To speed up the convergence of the coordinate descent algorithm, a Bernoulli sampling strategy for the coordinate selection is proposed in \cite{Dong2020}.

While \cite{Haghighatshoar2018,Chen2019a,Dong2020} assume single-cell systems, the use of the covariance approach for multi-cell systems is considered in \cite{ShaoUnsourced2020}, where a cooperative activity detection strategy that exploits the sparsity and similarity between neighboring BSs is proposed. This paper differs from \cite{ShaoUnsourced2020} in assuming the knowledge of channel large-scale fading, which is assumed to be unknown in \cite{ShaoUnsourced2020}. One of the main conclusions of this paper is that exploiting large-scale fading significantly impacts the activity detection performance in the multi-cell setting. In addition, this paper further considers the effect of finite-capacity constraints on the fronthaul links between the BSs and the CU. The covariance approach is also used in \cite{Ganesan2020} for the cell-free massive MIMO system. Although \cite{Ganesan2020} also assumes that the large-scale fading is available, \cite{Ganesan2020} exploits the large-scale fading information in a different way and we compare the performances numerically. Finally, we mention
the work \cite{Jiang2020a}, which studies the impact of the inter-cell interference on the covariance based activity detection. However, cooperation among the BSs is not considered in \cite{Jiang2020a}.

\subsection{Main Contributions}
This paper studies the sparse activity detection in a multi-cell massive MIMO system with C-RAN architecture by exploiting the channel large-scale fading information. The main contributions of this work are as follows:

\begin{itemize}
\item By incorporating the large-scale fading information into the covariance based method, this paper formulates the cooperative activity detection problem as an MLE problem and analyzes the performance limit of the MLE in the massive MIMO regime, where the number of antennas at each BS tends to infinity asymptotically. A numerical phase transition analysis is then carried out, which characterizes the required length of the signature sequences to ensure reliable activity detection for a given number of (active) devices.

\item The phase transition analysis demonstrates the importance of exploiting the large-scale fading information for activity detection in a cooperative multi-cell system. Numerically, the phase transition curve in the multi-cell scenario is shown to be similar to that in the single-cell scenario, which means that the required signature sequence length in a cooperative multi-cell system can be made to be comparable to that of a single-cell scenario when the number of antennas is sufficiently large.

\item To deal with the capacity limits of the fronthaul links between the BSs and the CU, this paper proposes a quantization scheme at the BSs that includes a preliminary activity detection and a uniform quantization of the detection results at the BSs for fronthaul transmission as the first stage. At the CU, the sample covariance matrices are reconstructed, based on which a final cooperative activity detection is performed. As compared to directly quantizing the entries of the sample covariance matrices at the BSs, the proposed method can substantially reduce the fronthaul rate requirements.
\end{itemize}

\subsection{Paper Organization and Notation}
The remainder of this paper is organized as follows. Section~\ref{sec.system} introduces the system model. Section~\ref{sec.single} considers the special single-cell case, and Section~\ref{sec.single.analysis} demonstrates how the large-scale fading information can be incorporated into the phase transition analysis. Section~\ref{sec.multi} considers the activity detection in the multi-cell case; its phase transition analysis is presented in Section~\ref{sec.multi.analysis}. Section~\ref{sec.compare} compares the phase transition analysis in the single-cell and multi-cell cases. Section~\ref{sec.quan} considers the quantization design accounting for capacity-limited fronthaul links. Numerical results are provided in Section~\ref{sec.simu}. Conclusions are drawn in Section~\ref{sec.conclusion}.

Throughout this paper, lower-case, boldface lower-case, and boldface upper-case letters denote scalars, vectors, and matrices, respectively. Calligraphy letters denote sets. Superscripts $(\cdot)^{H}$, $(\cdot)^{T}$, $(\cdot)^{*}$ denote conjugate transpose, transpose, conjugate, respectively. Further, $\mathbf{I}$ denotes identity matrix with appropriate dimensions, $\mathbb{E}[\cdot]$ denotes expectation, $|\cdot|$ denotes the determinant of a matrix or the cardinality of a set, $\mathrm{tr}(\mathbf{X})$ denotes the trace of $\mathbf{X}$, $\mathrm{diag}(x_1,\ldots,x_n)$ (or $\mathrm{diag}(\mathbf{X}_1,\ldots,\mathbf{X}_n)$) denotes a (block) diagonal matrix formed by $x_1,\ldots,x_n$ (or $\mathbf{X}_1,\ldots,\mathbf{X}_n$), $\|\mathbf{x}\|_1$ denotes the $\ell_1$ norm of $\mathbf{x}$.
We use $\triangleq$ to denote definition, $\odot$ for element-wise product, and $\otimes$ for Kronecker product.
A complex Gaussian distribution with mean $\boldsymbol\mu$ and covariance $\mathbf{\Sigma}$
is denoted by $\mathcal{CN}(\boldsymbol\mu, \mathbf{\Sigma})$.

\section{System model}
\label{sec.system}
Consider an uplink massive MIMO multi-cell system consisting of $B$ cells. Each cell contains one BS equipped with $M$ antennas. We assume a C-RAN architecture, in which all $B$ BSs are connected to a CU via fronthaul links such that the received signals can be collected and jointly processed at the CU for inter-cell interference mitigation. We assume that there are $N$ single-antenna devices in each cell but only $K\ll N$ devices are active during any coherence interval. Let $a_{bn}$ indicate the activity of device $n$ in cell $b$, i.e., $a_{bn}=1$ if the device is active and $a_{bn}=0$ otherwise. Let $g_{ibn}\mathbf{h}_{ibn}$ denote the channel between device $n$ in cell $b$ and BS $i$, where $\mathbf{h}_{ibn}\in\mathbb{C}^{M}$ is the Rayleigh fading component following $\mathcal{CN}(\mathbf{0},\mathbf{I})$, and $g_{ibn}$ is the large-scale fading coefficient including path-loss and shadowing. For device identification, each device is associated with a unique sequence $\mathbf{s}_{bn}\in\mathbb{C}^{L}$ with $L$ being the sequence length. In the uplink pilot stage, all active devices transmit their signature sequences as random access requests. Assuming that the sequences are transmitted synchronously, the received signal at BS $b$ can be expressed as
\begin{align}\label{eq.sys}
\mathbf{Y}_b&=\sum_{n=1}^{N}a_{bn}\mathbf{s}_{bn}g_{bbn}\mathbf{h}_{bbn}^T  + \sum_{j\neq b} \sum_{n=1}^N a_{jn}\mathbf{s}_{jn}g_{bjn}\mathbf{h}_{bjn}^T+\mathbf{W}_b\nonumber\\
&=\mathbf{S}_b\mathbf{A}_b\mathbf{G}^{\frac{1}{2}}_{bb}\mathbf{H}_{bb}+\sum_{j\neq b}\mathbf{S}_j\mathbf{A}_j\mathbf{G}^{\frac{1}{2}}_{bj}\mathbf{H}_{bj}+\mathbf{W}_b,
\end{align}
where $\mathbf{S}_j=[\mathbf{s}_{j1},\ldots,\mathbf{s}_{jN}]\in\mathbb{C}^{L\times N}$ is the signature sequence matrix of the devices in cell $j$, $\mathbf{A}_{j}=\operatorname{diag}\{a_{j1},\ldots,a_{jN}\}$ is a diagonal matrix that indicates the activity of the devices in cell $j$, $\mathbf{G}_{bj}=\operatorname{diag}\{g_{bj1}^2,\ldots,g_{bjN}^2\}$ contains the large-scale fading components between the devices in cell $j$ and BS $b$, $\mathbf{H}_{bj}=[\mathbf{h}_{bj1},\ldots,\mathbf{h}_{bjN}]^T\in\mathbb{C}^{N\times M}$ is the Rayleigh fading channel between the devices in cell $j$ and BS $b$, and $\mathbf{W}_b$ is the additive Gaussian noise that follows $\mathcal{CN}(\mathbf{0},\sigma_w^2\mathbf{I})$, where $\sigma_w^2$ is the variance of the background noise normalized by the transmit power for notational simplicity. Since both $\mathbf{A}_{j}$ and $\mathbf{G}_{bj}$ are diagonal, we can define a diagonal matrix $\boldsymbol\Gamma_{bj}\triangleq\mathbf{A}_j\mathbf{G}_{bj}\in \mathbb{R}^{N\times N}$ with diagonal entries $\gamma_{bjn}\triangleq a_{jn}g_{bjn}^2, \forall n$, and re-write \eqref{eq.sys} as
\begin{align}\label{eq.sys2}
\mathbf{Y}_b=\mathbf{S}_b\boldsymbol\Gamma^{\frac{1}{2}}_{bb}\mathbf{H}_{bb}+\sum_{j\neq b}\mathbf{S}_j\boldsymbol\Gamma^{\frac{1}{2}}_{bj}\mathbf{H}_{bj}+\mathbf{W}_b.
\end{align}
For notational simplicity, we also use $\mathbf{a}_{j}\triangleq[a_{j1},\ldots,a_{jN}]^T\in \mathbb{R}^N$ and $\boldsymbol\gamma_{bj}\triangleq[\gamma_{bj1},\ldots,\gamma_{bjN}]^T\in \mathbb{R}^N$ to denote the diagonal entries of $\mathbf{A}_{j}$ and $\boldsymbol\Gamma_{bj}$, respectively.

The aim is to detect the active devices in the system based on the received signals
$\mathbf{Y}_b, b=1,\cdots,B$ (or their quantized version, if the fronthaul links between
the BSs and the CU have finite-capacity constraint). This is equivalent to
estimating the activity indicator matrix $\mathbf{A}_b, \forall b$ in \eqref{eq.sys}, or alternatively the non-zero diagonal entries of $\boldsymbol\Gamma_{bb}, \forall b$ in \eqref{eq.sys2}, since the activity indicator is binary.
It can be shown that a sufficient statistic of this detection problem is the set of
sample covariance matrices of the received signals, defined as
\begin{align}\label{eq.sampCov}
\hat{\boldsymbol\Sigma}_b = \frac{1}{M}\mathbf{Y}_b\mathbf{Y}_b^H, \qquad b=1,\cdots, B.
\end{align}
So instead of using the received signals $\mathbf{Y}_b, \forall b$, we can design
device activity detection algorithms based on the above sample covariance matrices.
This approach is known as the covariance method.

In this paper, we assume that the large-scale fading components
$\mathbf{G}_{bj}, \forall b,j$ are known at the BS, and study the benefits of
exploiting the large-scale fadings in the covariance based device activity
detection.

\section{Covariance Based Device Activity Detection in Single-Cell Systems}
\label{sec.single}

We start by considering the special single-cell scenario ($B=1$) to show how
the large-scale fading information can be incorporated into the covariance based
method and the impact of knowing the large-scale fading on the phase transition
of the covariance based activity detection method.
To simplify the notation, we remove all the cell and BS indices in \eqref{eq.sys} and \eqref{eq.sys2}, and index the devices by $n$ only.

\subsection{Covariance Approach}
We briefly review the basic idea of the covariance approach for activity detection. From \eqref{eq.sys}, the received signal at the BS in the single-cell scenario can be written as
\begin{align}\label{eq.sys.single.cell}
\mathbf{Y}=\mathbf{S}\mathbf{A}\mathbf{G}^{\frac{1}{2}}\mathbf{H}+\mathbf{W}.
\end{align}
where $\mathbf{S}$ denotes the signature sequence matrix of the devices in a single-cell setup. The covariance approach treats the activity indicator matrix $\mathbf{A}$ and the large-scale fading matrix $\mathbf{G}$ as a set of deterministic parameters while treating the Rayleigh fading matrix $\mathbf{H}$ and noise $\mathbf{W}$ as random samples drawn from complex Gaussian distributions. Since the signature sequence matrix $\mathbf{S}$ is fixed, the covariance of the received signal $\mathbf{Y}$ can be computed as
\begin{align}\label{eq.trueCov}
\boldsymbol\Sigma &= \frac{1}{M}\mathbb{E}\left[\mathbf{Y}\mathbf{Y}^H\right]=\mathbf{S}\mathbf{A}\mathbf{G}\mathbf{S}^H+\sigma_w^2\mathbf{I}.
\end{align}
where the expectation is taken with respect to the Rayleigh fading components and the background noise.

To detect the active devices, we consider the MLE of $\mathbf{A}$ or $\boldsymbol\Gamma$, depending on whether the large-scale fading matrix $\mathbf{G}$ is known, based on $\mathbf{Y}$.
Notice from \eqref{eq.sys.single.cell} that the received signals at the $M$ antennas are i.i.d.\ due to the fact that the Rayleigh
fading components are i.i.d.\ over the antennas. Let $\mathbf{y}_{m}$ denote the received signal at the $m$-th antenna. We have that $\mathbf{y}_{m}$ follows $\mathcal{CN}(\mathbf{0},\boldsymbol\Sigma)$, where $\boldsymbol\Sigma$ is given in \eqref{eq.trueCov}.

If the large-scale fading matrix $\mathbf{G}$ is known, the likelihood function of
observing $\mathbf{Y}$ given $\mathbf{A}$ can be expressed as
\begin{align}\label{eq.likelihood}
p(\mathbf{Y}|\mathbf{a})&= \prod_{m=1}^Mp(\mathbf{y}_m|\mathbf{a})\nonumber\\
&=\frac{1}{|\pi\boldsymbol\Sigma|^M}\exp{\left(-\operatorname{tr}\left(M\boldsymbol\Sigma^{-1}\hat{\boldsymbol\Sigma}\right)\right)}, \end{align}
and it depends on the received signal $\mathbf{Y}$ via the sample covariance matrix $\hat{\boldsymbol\Sigma}$ only.
Thus, the maximization of $p(\mathbf{Y}|\mathbf{a})$ can be cast as the following minimization problem
\begin{subequations}\label{eq.prob1}
\begin{alignat}{2}\label{eq.prob1.1}
&\underset{\mathbf{a}}{\operatorname{minimize}}    &\quad& \log\left|\boldsymbol\Sigma\right|+ \operatorname{tr}\left(\boldsymbol\Sigma^{-1}\hat{\boldsymbol\Sigma}\right)\\
&\operatorname{subject\,to} &      &a_{n} \in \{0,1\}, \,\forall n.
\label{eq.prob1.3}
\end{alignat}
\end{subequations}
Problem \eqref{eq.prob1} is a challenging combinatorial problem due to the binary constraint \eqref{eq.prob1.3}. To make the problem \eqref{eq.prob1} more tractable, we can relax the constraint $a_{n} \in \{0,1\}$ to $a_{n} \in [0,1]$, and express the relaxed problem as
\begin{subequations}\label{eq.prob.single.lsf}
\begin{alignat}{2}\label{eq.prob.single.lsf.1}
&\underset{\mathbf{a}}{\operatorname{minimize}}    &\quad& \log\left|\boldsymbol\Sigma\right|+ \operatorname{tr}\left(\boldsymbol\Sigma^{-1}\hat{\boldsymbol\Sigma}\right)\\
&\operatorname{subject\,to} &      &a_{n} \in [0,1],\,\forall n.
\label{eq.prob.single.lsf.2}
\end{alignat}
\end{subequations}

The above formulation can be contrasted with the case of the large-scale fading matrix $\mathbf{G}$ being unknown, in which case
the likelihood function $p(\mathbf{Y}|\boldsymbol\gamma)$ remains a Gaussian distribution function, but the constraints are different. In this case,
the MLE of $\boldsymbol\Gamma$ can be formulated as \cite{Haghighatshoar2018}
\begin{subequations}\label{eq.prob.single.nolsf}
\begin{alignat}{2}\label{eq.prob.single.nolsf.1}
&\underset{\boldsymbol\gamma}{\operatorname{minimize}}    &\quad& \log\left|\boldsymbol\Sigma\right|+ \operatorname{tr}\left(\boldsymbol\Sigma^{-1}\hat{\boldsymbol\Sigma}\right)\\
&\operatorname{subject\,to} &      &\gamma_{n} \in [0,+\infty), \,\forall n,
\label{eq.prob.single.nolsf.2}
\end{alignat}
\end{subequations}
where the constraint \eqref{eq.prob.single.nolsf.2} is due to the fact that the large-scale fading components are non-negative.

\subsection{Coordinate Descent Algorithms}
The activity detection problems \eqref{eq.prob.single.lsf} and \eqref{eq.prob.single.nolsf} are in a similar form, so they can be solved using the same algorithmic framework.
Although the problems are not convex (because the log-determinant term is concave),
the coordinate descent algorithm that iteratively updates the coordinates of $\mathbf{a}$ and $\boldsymbol\gamma$ is quite effective for solving both \eqref{eq.prob.single.lsf} and \eqref{eq.prob.single.nolsf}.

For the case where the large-scale fading matrix $\mathbf{G}$ is known, the coordinate descent algorithm updates the current estimate $\hat{a}_{n}$ of $a_n$ by $\hat{a}_{n}\leftarrow\hat{a}_{n}+d$ with other $\hat{a}_{j}, j\neq n$ fixed at each iteration, and the resulting optimization problem to determine $d$ can be expressed as
\begin{alignat}{2}
&\underset{d}{\min}    &\quad& \log\left(1+dg_{n}\mathbf{s}_{n}^H\tilde{\boldsymbol\Sigma}^{-1}\mathbf{s}_{n}\right) -\frac{dg_{n}\mathbf{s}_{n}^H\tilde{\boldsymbol\Sigma}^{-1}\hat{\boldsymbol\Sigma}\tilde{\boldsymbol\Sigma}^{-1}\mathbf{s}_{n}}{1+dg_{n}\mathbf{s}_{n}^H\tilde{\boldsymbol\Sigma}^{-1}\mathbf{s}_{n}} \nonumber \\
&\operatorname{s.t.} &      &d \in [-\hat{a}_n,1-\hat{a}_n],
\label{eq.update.lsf}
\end{alignat}
where $\tilde{\boldsymbol\Sigma}=\sum_{n=1}^N\hat{a}_ng_n\mathbf{s}_n\mathbf{s}_n^H+\sigma_w^2\mathbf{I}$.
By taking the derivative of the objective function, a closed-form solution to \eqref{eq.update.lsf} can be derived as
\begin{align}\label{eq.soln.lsf}
d = \min\left\{\max\left\{\frac{\theta}{g_{n}},-\hat{a}_n\right\}, 1-\hat{a}_n\right\},
\end{align}
where
\begin{align}\label{eq.soln.theta} \theta=\frac{\mathbf{s}_{n}^H\tilde{\boldsymbol\Sigma}^{-1}\hat{\boldsymbol\Sigma}\tilde{\boldsymbol\Sigma}^{-1}\mathbf{s}_{n}
-\mathbf{s}_{n}^H\tilde{\boldsymbol\Sigma}^{-1}\mathbf{s}_{n}}
      {(\mathbf{s}_{n}^H\tilde{\boldsymbol\Sigma}^{-1}\mathbf{s}_{n})^2}.
\end{align}

For the case where the large-scale fading matrix $\mathbf{G}$ is unknown, the coordinate descent algorithm updates the current estimate $\hat{\gamma}_n$ by $\hat{\gamma}_{n}\leftarrow\hat{\gamma}_{n}+d$ with other $\hat{\gamma}_{n}, j\neq n$ fixed in a similar way as in \eqref{eq.update.lsf}, and $d$ can be determined in closed-form as \cite{Haghighatshoar2018}
\begin{align}\label{eq.soln.nolsf}
d = \max\left\{\theta,-\hat{\gamma}_n\right\},
\end{align}
where $\theta$ is given by \eqref{eq.soln.theta} with $\tilde{\boldsymbol\Sigma}=\sum_{n=1}^N\hat{\gamma}_n\mathbf{s}_n\mathbf{s}_n^H+\sigma_w^2\mathbf{I}$.

We observe from \eqref{eq.soln.lsf} and \eqref{eq.soln.nolsf} that the computational complexity in both cases is dominated by the matrix-vector multiplications and the matrix inversion in \eqref{eq.soln.theta}. Using rank-1 update for computing the matrix inversion, the complexity at each iteration can be reduced to $\mathcal{O}(L^2)$.

\section{Phase Transition Analysis of Covariance Based Detection in Single-Cell Case}
\label{sec.single.analysis}

We now present a performance analysis of the covariance based activity detection in the
single-cell scenario. The analysis is based on a phase transition in the massive MIMO
regime where $M$ tends to infinity.
Specifically, we answer the following question: Given $N$ potential devices in the cell and a set of non-orthogonal signature sequences $\mathbf{s}_n, \forall n$ of length $L$, how many active devices (i.e., $K$) can be reliably detected in the asymptotic massive MIMO regime?

Note that analyzing the performance of the coordinate descent algorithm for solving
\eqref{eq.prob.single.lsf} and \eqref{eq.prob.single.nolsf} is a challenging problem,
because the coordinate descent algorithm is not guaranteed to converge to the global
optimum solutions. Instead, we make an idealized assumption that the global optima of
\eqref{eq.prob.single.lsf} and \eqref{eq.prob.single.nolsf} can be obtained, and
focus on analyzing the performance of the globally optimal solutions.

For the case where the large-scale fading coefficients are not known, the phase
transition analysis has already been carried out in \cite{ChenZ2020}. In this
section, we first review the analysis in \cite{ChenZ2020}, then extend
the analysis to the case where the large-scale fadings are known, and finally
quantitatively compare the two cases.

\subsection{Case of Large-Scale Fading Not Known}

We first discuss the case where the large-scale fadings are unknown, and the device activity is obtained by solving \eqref{eq.prob.single.nolsf}. For notational clarity, let $\boldsymbol\gamma^0$ be the true value of $\boldsymbol\gamma$, and let $\hat{\boldsymbol\gamma}^{ML}$ be the optimal solution to \eqref{eq.prob.single.nolsf}. For device activity detection in the massive MIMO regime, we hope that
\begin{align}\label{eq.single.consist}
\hat{\boldsymbol\gamma}^{ML} \rightarrow \boldsymbol\gamma^0, \text{ as } M \rightarrow \infty,
\end{align}
so that the device activity can be reliably obtained from $\hat{\boldsymbol\gamma}^{ML}$. 
The condition \eqref{eq.single.consist} is also known as the consistency of the MLE.
Based on the estimation theory, the consistency requires that the true parameter is identifiable. However, for the device activity detection problem considered in this paper, the identifiability of
the true parameter $\boldsymbol\gamma^{0}$ is not obvious when the size of $\boldsymbol\gamma^0$ is much greater than the dimensions of the sample covariance matrix $\hat{\boldsymbol\Sigma}$, i.e., $N\gg L^2$. In such case, there are more parameters to estimate than the number of observations.

A precise characterization of the condition to ensure \eqref{eq.single.consist} is obtained in \cite{ChenZ2020} by analyzing the associated Fisher information matrix, which is defined as
\begin{align}\label{eq.fim.def.single}
[\mathbf{J}(\boldsymbol\gamma)]_{ij}=\mathbb{E}\left[\left(\frac{\partial \log p(\mathbf{Y}|\boldsymbol{\gamma})}{\partial \gamma_i}\right)\left(\frac{\partial \log p(\mathbf{Y}|\boldsymbol{\gamma})}{\partial \gamma_j}\right)\right],
\end{align}
where the expectation is taken with respect to $\mathbf{Y}$. By noticing that the likelihood function $p(\mathbf{Y}|\boldsymbol{\gamma})$ is a Gaussian distribution function, $\mathbf{J}(\boldsymbol\gamma)$ can be computed as\cite{ChenZ2020}
\begin{align}\label{eq.fim.single}
\mathbf{J}(\boldsymbol{\gamma}) = M\left(\mathbf{P}\odot \mathbf{P}^*\right),
\end{align}
where $\mathbf{P}\triangleq \mathbf{S}^H\left(\mathbf{S}\boldsymbol\Gamma\mathbf{S}^H+\sigma_w^2\mathbf{I}\right)^{-1}\mathbf{S}$.
The condition under which \eqref{eq.single.consist} is ensured can be stated as follows.

\begin{theorem}[\hspace{1sp}\cite{ChenZ2020}]\label{single.condition.nolsf}
Consider the MLE problem in \eqref{eq.prob.single.nolsf} for device activity detection with a given signature sequence matrix $\mathbf{S}$ and noise variance $\sigma_w^2$. Let $\hat{\boldsymbol{\gamma}}^{ML}$ be the solution to \eqref{eq.prob.single.nolsf}, and let $\boldsymbol{\gamma}^0$ be the true value of $\boldsymbol{\gamma}$ whose $N-K$ zero entries are indexed by $\mathcal{I}$, i.e.,
\begin{align}
\mathcal{I}\triangleq\{i\mid\gamma_i^0=0\}.
\end{align}
Define
\begin{align}
\mathcal{N}&\triangleq\{\mathbf{x}\mid \mathbf{x}^T\mathbf{J}(\boldsymbol{\gamma}^0)\mathbf{x}=0, \mathbf{x}\in \mathbb{R}^{N}\},\label{eq.single.subspace.nolsf}\\
\mathcal{C}&\triangleq\{\mathbf{x}\mid x_i\geq 0, i\in \mathcal{I}, \mathbf{x}\in \mathbb{R}^{N}\},\label{eq.single.cone.nolsf}
\end{align}
where $x_i$ is the $i$-th entry of $\mathbf{x}$. Then a necessary and sufficient condition for $\hat{\boldsymbol{\gamma}}^{ML}\rightarrow \boldsymbol{\gamma}^0$ as $M\rightarrow\infty$ is that the intersection of $\mathcal{N}$ and $\mathcal{C}$ is the zero vector, i.e., $\mathcal{N}\cap\mathcal{C}=\{\mathbf{0}\}$.
\end{theorem}

An interpretation of Theorem~\ref{single.condition.nolsf} is as follows.
\begin{itemize}
\item
$\mathcal{N}$ is the null space of $\mathbf{J}(\boldsymbol{\gamma}^0)$, which contains the directions starting from $\boldsymbol{\gamma}^0$ towards its neighborhood such that the likelihood function $p(\mathbf{Y}|\boldsymbol\gamma^0)$ stays unchanged, i.e.,
\begin{align}
p(\mathbf{Y}|\boldsymbol{\gamma}^0) = p(\mathbf{Y}|\boldsymbol{\gamma}^0+t\mathbf{x}),
\end{align}
for small positive $t$ and any $\mathbf{Y}$.
\item
$\mathcal{C}$ is a cone containing the directions starting from $\boldsymbol{\gamma}^0$ towards the feasible region $[0, \infty]^{N}$, i.e.,
\begin{align}
\boldsymbol{\gamma}^0+t\mathbf{x} \in [0,+\infty)^{N},
\end{align}
for small positive $t$.
\item $\mathcal{N}\cap\mathcal{C}=\{\mathbf{0}\}$ ensures that $\boldsymbol{\gamma}^0$ is uniquely identifiable around its neighborhood via the likelihood function $p(\mathbf{Y}|\boldsymbol{\gamma})$, which is necessary for $\hat{\boldsymbol{\gamma}}^{ML}\rightarrow\boldsymbol{\gamma}^0$. The sufficiency comes from the fact that the local and the global identifiability are equivalent for the device activity detection problem in \eqref{eq.prob.single.nolsf}.
\end{itemize}

Theorem~\ref{single.condition.nolsf} is useful in the sense that the condition $\mathcal{N}\cap\mathcal{C}=\{\mathbf{0}\}$ can be tested using linear programming \cite{ChenZ2020}, thus
providing a way to numerically identify the phase transition of the activity detection problem in the large $M$ limit.

It is worth mentioning that by plugging \eqref{eq.fim.single} into \eqref{eq.single.subspace.nolsf}, the expression of the subspace $\mathcal{N}$ can be simplified as \cite{ChenZ2020}
\begin{align}\label{eq.alt.subspace}
\mathcal{N}=\{\mathbf{x}\mid \widehat{\mathbf{S}}\mathbf{x}=\mathbf{0}, \mathbf{x}\in \mathbb{R}^{N}\},
\end{align}
where $\widehat{\mathbf{S}}$ is an $L^2\times N$ matrix formed by the column-wise Kronecker product of $\mathbf{S}^*$ and $\mathbf{S}$ as follows
\begin{align}\label{eq.col.wise.kron}
\widehat{\mathbf{S}}=[\mathbf{s}^{*}_{1}\otimes\mathbf{s}_{1}, \mathbf{s}^{*}_{2}\otimes\mathbf{s}_{2},\ldots, \mathbf{s}^{*}_{N}\otimes\mathbf{s}_{N}],
\end{align}
which indicates that $\widehat{\mathbf{S}}$ is determined by the signature sequence matrix only.

\subsection{Case of Known Large-Scale Fading}

We now extend the analysis of \cite{ChenZ2020} to the case where the large-scale fadings are known, and the device activity is obtained by solving \eqref{eq.prob.single.lsf}. Let $\mathbf{a}^0$ be the true value of $\mathbf{a}$, and let $\hat{\mathbf{a}}^{ML}$ be the solution to \eqref{eq.prob.single.lsf}. Similar to \eqref{eq.single.consist}, for reliable activity detection in the massive MIMO regime, we want to identify condition under which
$\hat{\mathbf{a}}^{ML}$ is consistent, i.e.,
\begin{align}\label{eq.single.consist.lsf}
\hat{\mathbf{a}}^{ML} \rightarrow \mathbf{a}^0, \text{ as } M \rightarrow \infty.
\end{align}
Following the approach used in the case with unknown large-scale fadings in the previous subsection, we analyze the Fisher information matrix associated with the MLE problem \eqref{eq.prob.single.lsf}, which is
\begin{align}\label{eq.single.fim.lsf}
[\mathbf{J}(\mathbf{a})]_{ij} &= \mathbb{E}\left[\left(\frac{\partial \log p(\mathbf{Y}|\mathbf{a})}{\partial a_i}\right)\left(\frac{\partial \log p(\mathbf{Y}|\mathbf{a})}{\partial a_j}\right)\right].
\end{align}
Note here a slight abuse of the notation for $\mathbf{J}(\cdot)$, but it is clear from the context that $\mathbf{J}(\mathbf{a})$ and $\mathbf{J}(\boldsymbol\gamma)$ denote two different matrices.
By plugging \eqref{eq.likelihood} into \eqref{eq.single.fim.lsf}, the Fisher information matrix can be computed as
\begin{align}\label{eq.single.fim.lsf.2}
\mathbf{J}(\mathbf{a})=M (\mathbf{Q}\odot\mathbf{Q}^*),
\end{align}
where $\mathbf{Q}=\mathbf{G}^{\frac{1}{2}}\mathbf{S}^H
(\mathbf{S}\mathbf{G}\mathbf{A}\mathbf{S}^H+\sigma_w^2\mathbf{I})^{-1}
\mathbf{S}\mathbf{G}^{\frac{1}{2}}$. Based on the Fisher information matrix, a necessary and sufficient condition to ensure consistency of $\hat{\mathbf{a}}^{ML}$ can be established.
\begin{theorem}\label{single.condition.lsf}
Consider the MLE problem in \eqref{eq.prob.single.lsf} for device activity detection with a given sequence matrix $\mathbf{S}$, noise variance $\sigma_w^2$, and a large-scale fading matrix $\mathbf{G}$. Let $\hat{\mathbf{a}}^{ML}$ be the solution to \eqref{eq.prob.single.lsf}, and let $\mathbf{a}^0$ be its true value whose $N-K$ zero entries are indexed by $\mathcal{I}$,
i.e.,
\begin{align}
\mathcal{I}\triangleq\{i\mid\gamma_i^0=0\}.
\end{align}
Define
\begin{align}
\mathcal{N}^\prime &\triangleq\{\mathbf{x}\mid \mathbf{x}^T\mathbf{J}(\mathbf{a}^0)\mathbf{x}=0, \mathbf{x}\in \mathbb{R}^{N}\},\label{eq.single.subspace.lsf}\\
\mathcal{C}^\prime&\triangleq\{\mathbf{x}\mid x_i\geq 0 \text{ if } i\in \mathcal{I}, x_i\leq 0 \text{ if } i\notin \mathcal{I}, \mathbf{x}\in \mathbb{R}^{N}\},\label{eq.single.cone.lsf}
\end{align}
where $x_i$ is the $i$-th entry of $\mathbf{x}$. Then a necessary and sufficient condition for $\hat{\mathbf{a}}^{ML}\rightarrow \mathbf{a}^0$ as $M\rightarrow\infty$ is that the intersection of $\mathcal{N}^\prime$ and $\mathcal{C}^\prime$ is the zero vector, i.e., $\mathcal{N}^\prime\cap\mathcal{C}^\prime=\{\mathbf{0}\}$.
\end{theorem}
\begin{IEEEproof}
The proof follows the same idea for proving Theorem~\ref{single.condition.nolsf} in \cite{ChenZ2020}. The only difference lies in the construction of $\mathcal{C}^\prime$, which is due to the constraints $0\leq a_i\leq 1, \forall i$, in \eqref{eq.prob.single.lsf.2} for the case where the large-scale fading information is exploited. Specifically, $\mathcal{C}^\prime$ defined in \eqref{eq.single.cone.lsf} ensures that for any $\mathbf{x}\in\mathcal{C}^\prime$ and small positive $t$, $\mathbf{a}^0+t\mathbf{x}$ is still feasible, i.e., $\mathbf{a}^0+t\mathbf{x}\in [0,1]^N$.
\end{IEEEproof}

Note that Theorem~\ref{single.condition.lsf} is established in a form similar to Theorem~\ref{single.condition.nolsf}, and thus the condition $\mathcal{N}^\prime\cap\mathcal{C}^\prime=\{\mathbf{0}\}$ in Theorem~\ref{single.condition.lsf} can also be interpreted similarly. Further, both conditions can be verified by linear programming. However, there are some important differences between these two conditions. First, the null spaces in \eqref{eq.single.subspace.lsf} and \eqref{eq.single.subspace.nolsf} are different, which is due to the fact that they correspond to different Fisher information matrices. To see this difference more clearly, we can plug \eqref{eq.single.fim.lsf.2} into \eqref{eq.single.subspace.lsf} to get a simplified expression of \eqref{eq.single.subspace.lsf} as follows
\begin{align}\label{eq.alt.subspace.lsf}
\mathcal{N}^\prime=\{\mathbf{x}\mid \widehat{\mathbf{S}}\mathbf{G}\mathbf{x}=\mathbf{0}, \mathbf{x}\in \mathbb{R}^{N}\}.
\end{align}
As compared to \eqref{eq.alt.subspace}, the null space $\mathcal{N}^\prime$ in \eqref{eq.alt.subspace.lsf} depends on not only the signature sequences but also the large-scale fading coefficients. Second, the cone $\mathcal{C}^\prime$ in \eqref{eq.single.cone.lsf} is much more constrained than that in \eqref{eq.single.cone.nolsf} due to the extra constraints $x_i\leq 0, i\notin \mathcal{I}$. These new constraints are caused by $a_i\leq 1, \forall i$, in \eqref{eq.prob.single.lsf.2}. 
The cone $\mathcal{C}^\prime$ in \eqref{eq.single.cone.lsf} can be seen as a collection of directions $\mathbf{x}\in\mathbb{R}^N$ starting from $\mathbf{a}^0$ such that $\mathbf{a}^0+t\mathbf{x}$ still lies in the feasible region, i.e.,
\begin{align}\label{eq.feasible}
\mathbf{a}^0+t\mathbf{x}\in [0, 1]^N,
\end{align}
for small positive $t$. It is necessary to require $x_i\leq 0, i\notin \mathcal{I}$ such that \eqref{eq.feasible} holds because $a_i^0=1, i\notin \mathcal{I}$.

It is worth noting that despite these differences, the two conditions in Theorem~\ref{single.condition.lsf} and Theorem~\ref{single.condition.nolsf} are closely related
as stated by the following proposition.
\begin{proposition}
$\mathcal{N}^\prime\cap\mathcal{C}^\prime\neq\{\mathbf{0}\}$ implies $\mathcal{N}\cap\mathcal{C}\neq\{\mathbf{0}\}$.
\end{proposition}
\begin{IEEEproof}
If there exits a non-zero vector $\mathbf{x}\in \mathcal{N}^\prime\cap\mathcal{C}^\prime$, we can then construct a real vector $\tilde{\mathbf{x}} = \mathbf{G}\mathbf{x}$ and it can be verified that $\tilde{\mathbf{x}}\in \mathcal{N}\cap\mathcal{C}$.
\end{IEEEproof}
This result agrees with the intuition that if we cannot get a consistent
estimation of the activity indicator $\mathbf{a}^0$ by exploiting the
large-scale fading coefficients, then we cannot get a consistent estimation of
$\boldsymbol\gamma$ without knowing the large-scale fadings. However, the
converse is not true, which means that we may still have $\hat{\mathbf{a}}^{ML}
\rightarrow \mathbf{a}^0$ in some cases even with $\hat{\boldsymbol\gamma}^{ML}
\nrightarrow \boldsymbol\gamma^0$ as $M$ tends to infinity.

There is an interesting symmetry property for the condition $\mathcal{N}^\prime\cap\mathcal{C}^\prime=\{\mathbf{0}\}$, due to the extra constraints $x_i\leq 0, i\notin \mathcal{I}$ in $\mathcal{C}^\prime$. Note that cone $\mathcal{C}^\prime$ in \eqref{eq.single.cone.lsf} is determined by $\mathcal{I}$.
Let $\mathcal{J}$ be the complement set of $\mathcal{I}$ with respect to $\{1,2,\ldots,N\}$, i.e.,
\begin{align}\label{eq.set.j}
\mathcal{J} = \{1,2,\ldots,N\} - \mathcal{I},
\end{align}
based on which we define cone $\bar{\mathcal{C}}^\prime$ similarly to \eqref{eq.single.cone.lsf} as follows
\begin{align}
\bar{\mathcal{C}}^\prime\triangleq\{\mathbf{x}\mid x_i\geq 0 \text{ if } i\in \mathcal{J}, x_i\leq 0 \text{ if } i \notin \mathcal{J}, \mathbf{x}\in \mathbb{R}^{N}\}.
\end{align}
We then have the following result.
\begin{proposition}\label{theo3}
$\mathcal{N}^\prime\cap\mathcal{C}^\prime=\{\mathbf{0}\}$ implies $\mathcal{N}^\prime\cap\bar{\mathcal{C}}^\prime=\{\mathbf{0}\}$, and vice versa.
\end{proposition}
\begin{IEEEproof}
This can be proved by noting that $\bar{\mathcal{C}}^\prime= - \mathcal{C}^\prime$ and $\mathcal{N}^\prime = - \mathcal{N}^\prime$. Here, the negative sign operates on every element of the set. The relationship $\bar{\mathcal{C}}^\prime= - \mathcal{C}^\prime$ holds because of \eqref{eq.set.j}, and $\mathcal{N}^\prime = - \mathcal{N}^\prime$ holds because $\mathcal{N}^\prime$ is a subspace.
\end{IEEEproof}

A consequence of Proposition~\ref{theo3} is as follows. Let us consider the following two device activity detection problems that differ in only the number of active devices: i) there are $K$ active devices which are specified by $\mathcal{I}$; and ii) there are $N-K$ active devices which are specified by $\mathcal{J}$. Proposition~\ref{theo3} states that in the large $M$ limit, if we can detect the $K$ active devices in $\mathcal{I}$, then we can also detect the $N-K$ active devices in $\mathcal{J}$, and vice versa. In other words, by exploiting the knowledge of large-scale fadings, the covariance based method is able to reliably detect the active devices not only when \emph{the active devices are sparse} (i.e., $\tfrac{K}{N} \ll 1$), but also when \emph{the inactive devices are sparse} (i.e., $\tfrac{N-K}{N}  \ll 1$). This is more powerful than the case without exploiting the knowledge of large-scale fadings.

It is important to state a caveat that
the phase transition analysis gives the performance limit of the covariance method as $M$ goes to infinity. In practice, $M$ is finite, and further the required values of $M$ to achieve a target detection error rate for different problems can be quite different. For the above problem, the number of antennas needed to achieve the phase transition limit when the inactive devices are sparse would actually have to be impractically large.
This is because the quality of the sample covariance matrices would degrade as the number of active devices becomes large.
In practice, the device activity detection problem always operates in the regime where the active devices are sparse. In this case, the performance for the case of knowing large-scale fading is only slightly better than not knowing large-scale fading in the single-cell setting.
Thus, the above result is mainly of theoretical interest.

\section{Covariance Based Cooperative Activity Detection in Multi-Cell Systems}
\label{sec.multi}

We now study the device activity detection problem in the multi-cell scenario, where the inter-cell interference becomes a limiting factor if each BS operates independently. To alleviate the inter-cell interference, we assume a C-RAN architecture and consider cooperative device activity detection at the CU. In this section, we assume infinite-capacity fronthaul links between the BSs and the CU such that the received signals at the BSs can be ideally collected at the CU.

\subsection{Limitation of Treating Interference as Noise}
\label{sec.tin}
We first discuss the impact of the inter-cell interference on the covariance based device activity detection if each BS simply treats the inter-cell interference as noise. We show that reliable activity detection may not be achieved using the covariance approach even in the massive MIMO regime.

Consider the case where the large-scale fadings are known. Based on \eqref{eq.sys}, the covariance of the received signal $\mathbf{Y}_b$ at BS $b$ can be expressed as
\begin{align}\label{eq.sys.multi.cell}
\boldsymbol\Sigma_b &= \frac{1}{M}\mathbb{E}\left[\mathbf{Y}_b\mathbf{Y}_b^H\right]\nonumber\\
&=\mathbf{S}_b\mathbf{A}_b\mathbf{G}_{bb}\mathbf{S}_b^H+\sum_{j\neq b}\mathbf{S}_j\mathbf{A}_j\mathbf{G}_{bj}\mathbf{S}_j^H+\sigma_w^2\mathbf{I},
\end{align}
where $\sum_{j\neq b}\mathbf{S}_j\mathbf{A}_j\mathbf{G}_{bj}\mathbf{S}_j^H$ is the covariance of the interference signals. BS $b$ aims to detect the device activity in cell $b$, i.e., $\mathbf{A}_b$, while treating $\sum_{j\neq b}\mathbf{S}_j\mathbf{A}_j\mathbf{G}_{bj}\mathbf{S}_j^H$ as noise.
As compared to \eqref{eq.trueCov} in the single-cell case,
we observe that the detection of $\mathbf{A}_b$ from $\boldsymbol\Sigma_b$ in \eqref{eq.sys.multi.cell} becomes much more challenging, because the covariance of the inter-cell interference
$\sum_{j\neq b}\mathbf{S}_j\mathbf{A}_j\mathbf{G}_{bj}\mathbf{S}_j^H$ depends on the device
activities in other cells which change over time and are difficult to track.
A straightforward way to deal with the time-varying interference covariance is to approximate $\sum_{j\neq b}\mathbf{S}_j\mathbf{A}_j\mathbf{G}_{bj}\mathbf{S}_j^H$ by its time average, which can be computed as
\begin{align}
\mathbb{E}\left[\sum_{j\neq b}\mathbf{S}_j\mathbf{A}_j\mathbf{G}_{bj}\mathbf{S}_j^H\right] = \frac{K}{N}\sum_{j\neq b}\mathbf{S}_j\mathbf{G}_{bj}\mathbf{S}_j^H,
\end{align}
where the expectation is taken with respect to $\mathbf{A}_j$ by noting that $K$ out of $N$ devices are active in each coherence interval. Besides using this time average, other possible methods include approximating the interference covariance by some structured matrix, e.g., a diagonal matrix or a scaled identity matrix \cite{Jiang2020a}. However, these methods inevitably introduce approximation errors in the modeling of the interference covariance matrix, which leads to severe performance loss when $M$ is large, as shown by simulation in Section~\ref{sec.simu}.
For the case where the large-scale fading components are unknown, the impact of the inter-cell interference is similar.

\subsection{Cooperative Detection with Unknown Large-Scale Fading}

Instead of treating inter-cell interference as noise at each BS, we now consider cooperative activity detection at the CU. We first consider the case where the large-scale fading coefficients are unknown, and the CU aims to estimate $\boldsymbol\Gamma_{bj}=\mathbf{A}_j\mathbf{G}_{bj}, \forall b,j \in \{1,\cdots,B\}$.

Similar to the single-cell case, we formulate the problem using the MLE of $\boldsymbol\Gamma_{bj}, \forall b,j \in \{1,\cdots,B\}$. Based on \eqref{eq.sys}, the received signal at the $m$-th antenna of BS $b$, denoted by $\mathbf{y}_{bm}$ follows $\mathcal{CN}(\mathbf{0},\boldsymbol\Sigma_b)$, where $\boldsymbol\Sigma_b$ is given in \eqref{eq.sys.multi.cell}.
Let $\underline{\boldsymbol{\gamma}}$ denote the collection of $\boldsymbol{\gamma}_{bj}, \forall b, j \in \{1,\cdots,B\}$.
The likelihood function of $\mathbf{Y}_{b}$ can be expressed as
\begin{align}\label{eq.likelihood.multi}
p(\mathbf{Y}_b|~\underline{\boldsymbol{\gamma}})=\frac{1}{|\pi\boldsymbol\Sigma_b|^M}\exp{\left(-\operatorname{tr}\left(M\boldsymbol\Sigma_b^{-1}\hat{\boldsymbol\Sigma}_b\right)\right)}.
\end{align}
Note that given $\underline{\boldsymbol\gamma}$, the received signals $\mathbf{Y}_{b}$ are independent because of the i.i.d.\ Rayleigh fading channels. The likelihood function of
$\mathbf{Y}_{1}, \cdots, \mathbf{Y}_{B}$ can then be written as
\begin{align}\label{eq.likelihoodall.multi}
p(\mathbf{Y}_1,\ldots,\mathbf{Y}_B|~\underline{\boldsymbol{\gamma}})= \prod_{b=1}^B p(\mathbf{Y}_b|~\underline{\boldsymbol{\gamma}}),
\end{align}
and the maximization of the log-likelihood function can be cast as a minimization problem as
\begin{subequations}\label{eq.prob.multi.nolsf}
\begin{alignat}{2}\label{eq.prob.multi.nolsf.1}
&\underset{\underline{\boldsymbol\gamma}}{\operatorname{minimize}}    &\quad& \sum_{b=1}^B\left(\log\left|\boldsymbol\Sigma_b\right|+ \operatorname{tr}\left(\boldsymbol\Sigma_b^{-1}\hat{\boldsymbol\Sigma}_b\right)\right)\\
&\operatorname{subject\,to} &      &\gamma_{bjn} \in [0,\infty), \,\forall b,j,n,
\label{eq.prob.multi.nolsf.2}
\end{alignat}
\end{subequations}
whose solution depends on the received signals at the BSs via the sample covariance matrices $\hat{\boldsymbol\Sigma}_b, b=1,\cdots,B$. Note that the above problem is very similar to the single-cell case, and it can also be solved using the coordinate descent algorithm with closed-form solutions.

However, it turns out that the detection performance of \eqref{eq.prob.multi.nolsf} is not much better than treating-interference-as-noise.
The main reason is that the number of unknown parameters to be estimated in \eqref{eq.prob.multi.nolsf} is $NB^2$, which is much larger than the number of the devices, $NB$.
In other words, the requirement to estimate the large-scale fadings $\boldsymbol\Gamma_{bj}$ from all the active devices to all the BSs is significantly more difficult than the single-cell case. 

It is worth noting that \eqref{eq.prob.multi.nolsf} is not the best possible problem formulation, because it does not take into consideration the interdependencies among $\boldsymbol\Gamma_{bj}$, which can in theory be exploited.
Such interdependencies come from two facts:
i) $\boldsymbol\Gamma_{bj}=\mathbf{A}_j\mathbf{G}_{bj}$ across $b=1,\cdots,B$
depend on the same activity indicator $\mathbf{A}_j$;
and ii) $\mathbf{G}_{bj}$ should satisfy certain constraints imposed by the geometry of the locations of the active devices and the locations of the BSs in the network.
But, it is quite challenging to incorporate such interdependencies into the detection algorithm;
see Section~\ref{sec.simu}.

\subsection{Cooperative Detection with Known Large-Scale Fading}
\label{sec.detection.multi}

The way to substantially improve detection performance is to gain knowledge of large-scale fadings and to exploit such knowledge in
cooperative activity detection at the CU.
When $\mathbf{G}_{bj}, \forall b,j$ are available, the CU can focus on detecting the device activities, i.e., $\mathbf{A}_b, \forall b$, without having to estimate $\mathbf{G}_{bj}$. In this case, the likelihood function $p(\mathbf{Y}_b|\mathbf{a}_1,\ldots,\mathbf{a}_B)$ remains a Gaussian distribution with covariance matrix $\boldsymbol\Sigma_b$. Further, $\mathbf{Y}_{b}$ for $b=1,\cdots,B$ are independent given $\mathbf{a}_1,\ldots,\mathbf{a}_B$.
Therefore, the maximization of the likelihood function $p(\mathbf{Y}_1,\ldots,\mathbf{Y}_B|\mathbf{a}_1,\ldots,\mathbf{a}_B)$ can be cast as the following minimization problem
\begin{subequations}\label{eq.prob1.multi}
\begin{alignat}{2}\label{eq.prob1.1.multi}
&\underset{\mathbf{a}_{1}, \cdots, \mathbf{a}_B}{\operatorname{minimize}}    &\quad& \sum_{b=1}^B\left(\log\left|\boldsymbol\Sigma_b\right|+ \operatorname{tr}\left(\boldsymbol\Sigma_b^{-1}\hat{\boldsymbol\Sigma}_b\right)\right)\\
&\operatorname{subject\,to} &      &a_{bn} \in \{0,1\}, \forall b,n
\label{eq.prob1.3.multi},
\end{alignat}
\end{subequations}
To deal with the binary constraint \eqref{eq.prob1.3.multi}, we consider a relaxed version of the problem as
\begin{subequations}\label{eq.prob2.multi}
\begin{alignat}{2}\label{eq.prob1.2.multi}
&\underset{\mathbf{a}_{b}, \cdots, \mathbf{a}_B}{\operatorname{minimize}}    &\quad& \sum_{b=1}^B\left(\log\left|\boldsymbol\Sigma_b\right|+ \operatorname{tr}\left(\boldsymbol\Sigma_b^{-1}\hat{\boldsymbol\Sigma}_b\right)\right)\\
&\operatorname{subject\,to} &      &a_{bn} \in [0,1], \,\forall b,n.
\label{eq.prob2.3.multi}
\end{alignat}
\end{subequations}
For the cooperative activity detection, we also use the coordinate descent framework to alternatively update the entries of $\mathbf{a}_{b}, \forall b=1,\cdots,B$. At each iteration, we update the estimate $\hat{a}_{bn}$ by $\hat{a}_{bn}\leftarrow\hat{a}_{bn}+d$ while keeping other activity indicators fixed. The optimization problem to determine  $d$ can be expressed as
\begin{subequations}\label{eq.prob4}
\begin{alignat}{2}\label{eq.prob4.1}
&\underset{d}{\operatorname{minimize}}    &\quad& \sum_{j=1}^B \log\left(1+dg_{jbn}\mathbf{s}_{bn}^H\tilde{\boldsymbol\Sigma}_j^{-1}\mathbf{s}_{bn}\right)\nonumber\\
&&&\qquad\qquad~~~-\frac{dg_{jbn}\mathbf{s}_{bn}^H\tilde{\boldsymbol\Sigma}_j^{-1}\hat{\boldsymbol\Sigma}_j\tilde{\boldsymbol\Sigma}_j^{-1}\mathbf{s}_{bn}}{1+dg_{jbn}\mathbf{s}_{bn}^H\tilde{\boldsymbol\Sigma}_j^{-1}\mathbf{s}_{bn}} \\
&\operatorname{subject\,to} &      &d \in [-\hat{a}_{bn},1-\hat{a}_{bn}].
\label{eq.prob4.3}
\end{alignat}
\end{subequations}
Note that unlike the problem \eqref{eq.update.lsf}, which admits a closed-form solution, the closed-form solution for the problem \eqref{eq.prob4} cannot be easily obtained because the derivative of the objective function involves a polynomial function whose degrees depend on $B$. However, since $d$ is a scalar, we can use a one-dimensional search to obtain the optimal $d$ or we can use a root-finding algorithm to identify the roots of the polynomial function and to determine the minimizer of \eqref{eq.prob4}.

\section{Phase Transition Analysis of Covariance Based Detection in Multi-Cell Case}
\label{sec.multi.analysis}

We now analyze the performance of the covariance based cooperative activity detection in \eqref{eq.prob2.multi} for multi-cell systems assuming that the large-scale fadings are known. Following the analysis in the single-cell case, we assume that \eqref{eq.prob2.multi} can be solved to global optimality, and focus on the phase transition analysis as $M$ tends to infinity.

Let $\hat{\mathbf{a}}_{b}^{ML}$ denote the solution to \eqref{eq.prob2.multi}, and let $\mathbf{a}_{b}^0$ be the true value of device activities for $b=1,\cdots,B$. We aim to identify the condition under which the estimates are consistent, i.e.,
\begin{align}\label{eq.multi.consist.lsf}
\hat{\mathbf{a}}_b^{ML} \rightarrow \mathbf{a}_b^0, \text{ as } M \rightarrow \infty,\, \forall b=1,\cdots,B.
\end{align}
Again, the Fisher information matrix plays an essential role in the analysis. We use the notation $\underline{\mathbf{a}}=[\mathbf{a}_{1}^T,\ldots,\mathbf{a}_{B}^T]^T\in\mathbb{R}^{BN}$ with the estimated and true values of $\underline{\mathbf{a}}$ denoted as $\underline{\hat{\mathbf{a}}}^{ML}$ and
$\underline{\mathbf{a}}^0$, respectively. Let
$\underline{\mathbf{A}}=\operatorname{diag}\left(\mathbf{A}_{1},\ldots,\mathbf{A}_{B}\right)\in\mathbb{R}^{BN\times BN}$,
$\mathbf{G}_b=\operatorname{diag}\left(\mathbf{G}_{b1},\ldots,\mathbf{G}_{bB}\right)\in\mathbb{R}^{BN\times BN}$, and $\underline{\mathbf{S}}=\left[\mathbf{S}_{1},\ldots,\mathbf{S}_{B}\right]\in\mathbb{C}^{L\times BN}$. Based on the likelihood function $p(\mathbf{Y}_b|\underline{\mathbf{a}})$, the Fisher information matrix of $\underline{\mathbf{a}}$ can be computed as
\begin{align}\label{eq.fim}
\mathbf{J}(\underline{\mathbf{a}}) = M\sum_{b=1}^B \left(\mathbf{Q}_b\odot\mathbf{Q}_b^*\right),
\end{align}
where $\mathbf{Q}_b=\mathbf{G}^{\frac{1}{2}}_b\underline{\mathbf{S}}^H
(\underline{\mathbf{S}}\mathbf{G}_b\underline{\mathbf{A}}\hspace{1pt}\underline{\mathbf{S}}^H+\sigma_w^2\mathbf{I})
\underline{\mathbf{S}}\mathbf{G}^{\frac{1}{2}}_b$. Based on the Fisher information matrix, the condition for the consistency of MLE can be established as follows:
\begin{theorem}\label{th.main.multi}
Consider the MLE problem in \eqref{eq.prob2.multi} for device activity detection with a given signature sequence matrix $\underline{\mathbf{S}}\in\mathbb{C}^{L \times BN}$ and noise variance $\sigma_w^2$. Let $\underline{\hat{\mathbf{a}}}^{ML}$ be the solution of \eqref{eq.prob2.multi}, and let $\underline{\mathbf{a}}^0$ be its true value whose $B(N-K)$ zero entries are indexed by $\underline{\mathcal{I}}$, i.e.,
\begin{align}
\underline{\mathcal{I}}\triangleq\{i\mid \underline{a}_i^0=0\},
\end{align}
where
$\underline{a}_i^0$ being the $i$-th entry of $\underline{\mathbf{a}}^0$. Define
\begin{align}
\mathcal{N}^{\prime\prime}&\triangleq\{\mathbf{x}\mid \mathbf{x}^T\mathbf{J}(\mathbf{a}^0)\mathbf{x}=0, \mathbf{x}\in \mathbb{R}^{BN}\},\label{eq.subspace}\\
\mathcal{C}^{\prime\prime}&\triangleq\{\mathbf{x}\mid x_i\geq 0 \text{ if } i\in \underline{\mathcal{I}}, x_i\leq 0 \text{ if } i \notin \underline{\mathcal{I}}, \mathbf{x}\in \mathbb{R}^{BN}\},\label{eq.cone}
\end{align}
where $x_i$ is the $i$-th entry of $\mathbf{x}$.
Then a necessary and sufficient condition for $\underline{\hat{\mathbf{a}}}^{ML}\rightarrow \underline{\mathbf{a}}^0$ as $M\rightarrow\infty$ is that the intersection of $\mathcal{N}^{\prime\prime}$ and $\mathcal{C}^{\prime\prime}$ is the zero vector, i.e., $\mathcal{N}^{\prime\prime}\cap\mathcal{C}^{\prime\prime}=\{\mathbf{0}\}$.
\end{theorem}
\begin{IEEEproof}
The proof follows the same line as in the proof of Theorem~\ref{single.condition.lsf} except for the difference in dimensionality.
\end{IEEEproof}

By plugging \eqref{eq.fim} into \eqref{eq.subspace} and by noticing that $\mathbf{Q}_b\odot\mathbf{Q}_b^*$ is positive semi-definite, \eqref{eq.subspace} can be re-written as $\mathcal{N}^{\prime\prime}=\{\mathbf{x}\mid \mathbf{x}^T(\mathbf{Q}_b\odot\mathbf{Q}_b^*)\mathbf{x}=0, \forall b, \mathbf{x}\in \mathbb{R}^{BN}\}$. Similar to \eqref{eq.alt.subspace.lsf}, the expression of $\mathcal{N}^{\prime\prime}$ can be further simplified as
\begin{align}\label{eq.cone2}
\mathcal{N}^{\prime\prime}=
\{\mathbf{x}\mid
\tilde{\mathbf{S}}\mathbf{G}_b\mathbf{x}=0, \forall b,\mathbf{x}\in \mathbb{R}^{BN}
\},
\end{align}
where $\tilde{\mathbf{S}}=[\mathbf{s}_{11}^*\otimes \mathbf{s}_{11},\ldots ,\mathbf{s}_{BN}^*\otimes \mathbf{s}_{BN}]\in\mathbb{C}^{L^2\times BN}$ is the column-wise Kronecker product of $\underline{\mathbf{S}}^*$ and $\underline{\mathbf{S}}$.
We observe from \eqref{eq.cone2} that $\mathcal{N}^{\prime\prime}$ is solely determined by the signature sequence matrix $\underline{\mathbf{S}}$ and the large-scale fading matrices $\mathbf{G}_b, \forall b$. 

Just as in the single-cell case, the condition $\mathcal{N}^{\prime\prime}\cap\mathcal{C}^{\prime\prime}=\{\mathbf{0}\}$ can be verified by linear programming under fixed values of $N, L, K$ and fixed $\mathbf{S}$ and $\mathbf{G}_b$.
This linear programming problem is explicitly written out in Section~\ref{sec.compare}.

As a further remark, $\mathcal{N}^{\prime\prime}\cap\mathcal{C}^{\prime\prime}=\{\mathbf{0}\}$ also has a symmetric property, similar to the single-cell case with known large-scale fadings. Let $\underline{\mathcal{J}}$ be the complement set of $\underline{\mathcal{I}}$ with respect to $\{1,2,\ldots,BN\}$. Define
\begin{align}
\bar{\mathcal{C}}^{\prime\prime}\triangleq\{\mathbf{x}\mid x_i\geq 0 \text{ if } i\in \underline{\mathcal{J}}, x_i\leq 0 \text{ if } i \notin \underline{\mathcal{J}}, \mathbf{x}\in \mathbb{R}^{BN}\}.
\end{align}
Then we have the following result.
\begin{proposition}\label{theo5}
$\mathcal{N}^{\prime\prime}\cap\mathcal{C}^{\prime\prime}=\{\mathbf{0}\}$ implies $\mathcal{N}^{\prime\prime}\cap\bar{\mathcal{C}}^{\prime\prime}=\{\mathbf{0}\}$, and vice versa.
\end{proposition}

The above result indicates that by exploiting the large-scale fading, the phase transition curves in the multi-cell and single-cell scenarios are similarly shaped.

\section{Comparison of Phase Transitions of Multi-Cell vs Single-Cell Systems}
\label{sec.compare}
We have seen that with known large-scale fadings, the phase transition curves of the covariance method in the multi-cell and single-cell cases are characterized by $\mathcal{N}^{\prime\prime}\cap\mathcal{C}^{\prime\prime}=\{\mathbf{0}\}$ and $\mathcal{N}^{\prime}\cap\mathcal{C}^{\prime}=\{\mathbf{0}\}$, respectively. 
We now explicitly compare these two via linear programming.

First, consider the multi-cell case, where we need $\mathcal{N}^{\prime\prime}\cap\mathcal{C}^{\prime\prime}=\{\mathbf{0}\}$. Note that $\tilde{\mathbf{S}}$ in \eqref{eq.cone2} is complex while $\mathcal{N}^{\prime\prime}$ is a real subspace, we need to separate the real and imaginary parts of $\tilde{\mathbf{S}}$. Let $\mathbf{r}_i^T\in \mathbb{C}^{1\times BN}$ be the $i$-th row of the signature matrix $\underline{\mathbf{S}}$. Since $\tilde{\mathbf{S}}$ is the column-wise Kronecker product of $\underline{\mathbf{S}}^*$ and $\underline{\mathbf{S}}$, the real and imaginary parts of rows of $\tilde{\mathbf{S}}$ can be represented by the following two sets of row vectors:
\begin{equation}
\left\{\operatorname{Re}(\mathbf{r}_i^T)\odot \operatorname{Re}(\mathbf{r}_j^T)+\operatorname{Im}(\mathbf{r}_i^T)\odot \operatorname{Im}(\mathbf{r}_j^T), 1\leq i\leq j\leq L\right\}
\label{eq.set_one}
\end{equation}
and
\begin{equation}
\left\{\operatorname{Re}(\mathbf{r}_i^T)\odot \operatorname{Im}(\mathbf{r}_j^T)-\operatorname{Im}(\mathbf{r}_i^T)\odot \operatorname{Re}(\mathbf{r}_j^T), 1\leq i< j\leq L\right\}.
\label{eq.set_two}
\end{equation}
These two sets consist of $L^2$ vectors in $\mathbb{R}^{1\times BN}$.
Let $\mathbf{D}\in \mathbb{R}^{L^2\times BN}$ be the matrix formed by all $L^2$ row vectors from the two sets. Furthermore, let $\tilde{\mathbf{D}}=[(\mathbf{D}\mathbf{G}_1)^T,\ldots,(\mathbf{D}\mathbf{G}_B)^T]^T\in \mathbb{R}^{BL^2\times BN}$. Then the condition $\mathcal{N}^{\prime\prime}\cap\mathcal{C}^{\prime\prime}=\{\mathbf{0}\}$ is equivalent to the infeasibility of the following linear programming problem
\begin{subequations}
\begin{alignat}{2}
&\quad \operatorname{find} &\quad& \mathbf{x}\label{eq.apd.lp.alt.1}\\
&\operatorname{subject\,to}&      & \tilde{\mathbf{D}}\mathbf{x} = \mathbf{0},\label{eq.apd.lp.alt.2}\\
& & &{x}_i\geq 0, i\in \underline{\mathcal{I}},\label{eq.apd.lp.alt.4}\\
& & &{x}_i\leq 0, i\notin \underline{\mathcal{I}},\label{eq.apd.lp.alt.5}\\
& & &\sum_{i\in \underline{\mathcal{I}}}x_i - \sum_{i\notin \underline{\mathcal{I}}}x_i = 1,\label{eq.apd.lp.alt.3}
\end{alignat}\label{eq.apd.lp.alt}%
\end{subequations}
where $\mathbf{x}\in\mathbb{R}^{BN}$, and the constraint \eqref{eq.apd.lp.alt.3} guarantees that $\mathbf{x}$ is non-zero. We observe from \eqref{eq.apd.lp.alt.2} that $\mathbf{x}$ should lie in a $(BN-BL^2)$ dimensional null space of $\tilde{\mathbf{D}}$. This is a random subspace due to the randomness of the signature matrix and the large-scale fadings. Notice here that knowing the large-scale fadings is crucial since the definition of this subspace depends on $\mathbf{G}_1,\mathbf{G}_2,\ldots,\mathbf{G}_B$.

Next consider the single-cell case corresponding to the condition
$\mathcal{N}^{\prime}\cap\mathcal{C}^{\prime}=\{\mathbf{0}\}$.
The real and imaginary parts of rows of $\widehat{\mathbf{S}}$ in \eqref{eq.alt.subspace.lsf} can be expressed by $L^2$ vectors in
$\mathbb{R}^{1\times N}$ similar to \eqref{eq.set_one} and \eqref{eq.set_two},
and these $L^2$ vectors can form an $L^2\times N$ matrix $\mathbf{C}$. Let
$\tilde{\mathbf{C}}=\mathbf{C}\mathbf{G}\in \mathbb{R}^{L^2\times N}$. Then the
condition $\mathcal{N}^{\prime}\cap\mathcal{C}^{\prime}=\{\mathbf{0}\}$ is
equivalent to the infeasibility of the following linear programming problem
\begin{subequations}
\begin{alignat}{2}
&\quad \operatorname{find} &\quad& \mathbf{x}\label{eq.apd.lp.alt.single.1}\\
&\operatorname{subject\,to}&      & \tilde{\mathbf{C}}\mathbf{x} = \mathbf{0},\label{eq.apd.lp.alt.single.2}\\
& & &{x}_i\geq 0, i\in\mathcal{I},\label{eq.apd.lp.alt.single.4}\\
& & &{x}_i\leq 0, i\notin\mathcal{I},\label{eq.apd.lp.alt.single.5}\\
& & &\sum_{i\in \mathcal{I}}x_i - \sum_{i\notin \mathcal{I}}x_i = 1,\label{eq.apd.lp.alt.single.3}
\end{alignat}\label{eq.apd.lp.alt.single}%
\end{subequations}
where $\mathbf{x}\in\mathbb{R}^{N}$.

We observe that problems \eqref{eq.apd.lp.alt} and \eqref{eq.apd.lp.alt.single} are in the same form, except the difference in dimensionality. The problem \eqref{eq.apd.lp.alt} aims to find a non-zero vector in a $(BN-BL^2)$ dimensional subspace with its $(BN-BK)$ entries being non-negative and the other $BK$ entries being non-positive, whereas the problem \eqref{eq.apd.lp.alt.single} aims to find a non-zero vector in an $(N-L^2)$ dimensional subspace with its $(N-K)$ entries being non-negative and the other $K$ entries being non-positive.

The crucial fact is that the dimension of the subspace and the number of the constraints simply scale by a factor of $B$ when we go from the single-cell to the multi-cell case.
Since the phase transitions depend on the ratio of these dimensions, we expect the phase transition curve for the multi-cell to be similar to that of the single-cell cases.
We numerically compare the two in Section~\ref{sec.simu}, and observe that the curves in fact closely match each other.

Note that when the large-scale fadings are not known, the phase transitions for the single-cell and for the multi-cell cases would be completely different.
This is because the corresponding null spaces have the dimension $N-L^2$ in the single-cell case, and $BN-L^2$ in the multi-cell case; they are not scaled versions of each other.

\section{Quantization Design for Covariance Based Cooperative Detection}
\label{sec.quan}
So far, we have assumed that the received signals (or the sample covariance matrices) at the BSs can be made available exactly to the CU for joint detection. In practice, the BSs and the CU are connected by capacity-limited fronthaul links, thus the signals at the BSs need to be quantized before they can be forwarded to the CU. The quantization error impairs the cooperative activity detection at the CU. In this section, we study the design of quantization schemes for cooperative activity detection under capacity-limited fronthauls. We consider cooperative activity detection that exploits the large-scale fading information.

The design of the optimal quantization scheme for cooperative activity detection is challenging due to the fact that the detection result obtained from \eqref{eq.prob2.multi} is a complicated function of the sample covariance matrices. In this section, we consider the following two heuristic quantization strategies: 1) Each BS computes and quantizes entries of its local sample covariance matrix to forward to the CU; 2) Each BS performs a preliminary activity detection based on the sample covariance matrix and quantizes the detection results for transmission to the CU, where the sample covariance matrices are approximately reconstructed and the cooperative activity detection is then performed.
We show that this novel second strategy has a significantly better performance.

\subsection{Quantization of Sample Covariance Matrix}

Since the cooperative activity detection at the CU, i.e., \eqref{eq.prob2.multi},
depends on the received signals $\mathbf{Y}_b$ via the sample covariance matrices
$\hat{\boldsymbol\Sigma}_b$, instead of quantizing $\mathbf{Y}_b$ of size $L\times M$,
a more efficient way at large $M$ is to quantize the $L\times L$ matrix
$\hat{\boldsymbol\Sigma}_b$. We use a uniform scalar quantizer as follows
\begin{align}
\hat{\boldsymbol\Sigma}_b^Q = \mathcal{Q}(\hat{\boldsymbol\Sigma}_b),\, \forall b,
\end{align}
where $\mathcal{Q}(\cdot)$ is a real-valued uniform scalar quantization function applied to the real and imaginary parts separately of each entry of $\hat{\boldsymbol\Sigma}_b$. We assume that each real or imaginary part is quantized with $2^{R_s}$ levels, where $R_s$ is the number of quantization bits per scalar. Since $\hat{\boldsymbol\Sigma}_b$ is a Hermitian matrix, the BS only needs to transmit the upper (or lower) triangular part of $\hat{\boldsymbol\Sigma}_b$ to the CU, which includes $L^2$ real-valued scalars. Thus the overall number of bits is $R_sL^2$.
Note that
the dynamic range of the elements of $\hat{\boldsymbol\Sigma}_b$ can be large due to the randomness of the channels and the device activity. To achieve a small quantization error, a large $R_s$ is usually needed.

\subsection{Quantization Based on Preliminary Detection}

In this subsection, we propose a novel quantization strategy in which each BS
performs a local preliminary detection of nearby devices based on its local
sample covariance matrix, then sends quantized detection results to the CU.
This is more efficient than directly quantizing the sample covariance matrix.
Specifically, each BS $b$ performs a local detection to estimate the activity
indicators $a_{jn}$ based on $\hat{\boldsymbol\Sigma}_b$ by solving the
following problem,
\begin{subequations}\label{eq.prob.multi.nolsf.q}
\begin{alignat}{3}\label{eq.prob.multi.nolsf.1.q}
&\underset{\{\mathbf{a}_{j}, \forall j\}}{\operatorname{minimize}}    &\quad& \log\left|\boldsymbol\Sigma_b\right|+ \operatorname{tr}\left(\boldsymbol\Sigma_b^{-1}\hat{\boldsymbol\Sigma}_b\right)\\
&\operatorname{subject\,to} &      &a_{jn} \in [0,1], \,\forall j,n,
\label{eq.prob.multi.nolsf.2.q}
\end{alignat}
\end{subequations}
which corresponds to maximizing the log-likelihood function $p(\mathbf{Y}_b|\mathbf{a}_1,\ldots,\mathbf{a}_B)$.
In problem formulation \eqref{eq.prob.multi.nolsf.q}, we assume that BS $b$ detects all the devices in the $B$ cells for simplicity. However, due to computational complexity and communication overhead, each BS $b$ can typically only detect activities of limited number of devices in practice. For example, we can consider a scenario in which each BS only aims to detect the activity pattern of devices in its own cell and the devices in the immediate neighboring cells. This is a reasonable scenario since the interference from other non-neighboring cells is typically weak and can be incorporated into the noised term.
We further note that to perform the optimization in \eqref{eq.prob.multi.nolsf.q},
all the sequence matrices $\mathbf{S}_j, \forall j$ need to be known at each BS $b$, which can be
accomplished by information exchange via the CU.

Similar to problem \eqref{eq.prob.single.lsf} for the single-cell case, problem \eqref{eq.prob.multi.nolsf.q} can be solved efficiently using the coordinate descent algorithm with closed-form solutions. We denote the results of \eqref{eq.prob.multi.nolsf.q} by $\hat{a}_{jn}^{b}, \forall j,n$, where superscript $b$ indicates that these estimates are obtained at BS $b$. We then quantize $\hat{a}_{jn}^{b}$ using a uniform scalar quantizer as follows
\begin{align}
\bar{a}_{jn}^{b} = \mathcal{Q}(\hat{a}_{jn}^{b}), \, \forall j,n,
\end{align}
where $\mathcal{Q}(\cdot)$ is the real-valued quantization function with $2^{R_a}$ levels. Since $\hat{a}_{jn}^{b}\in [0,1]$, the dynamic range of $\hat{a}_{jn}^{b}$ is much smaller than that of the entries of the sample covariance matrix, and therefore the value of $R_a$ could be much less than $R_s$ in the quantization of the sample covariance matrix.
Since the number of estimated activity indicators is $NB$, the overall number of bits is $R_aNB$ using fixed-length codewords to represent quantization levels.

We point out that the number of bits can be substantially reduced by using variable-length lossless data compression such as Huffman coding. This is because most of the indicators $\hat{a}_{jn}^{b}$ are zero or one, due to the fact that the device activities are sparse. Therefore, the overall number of bits needed to be forwarded to the CU can be much smaller than that in the quantization of the sample covariance.

Once the quantized activity indicators $\bar{a}_{jn}^{b}, \forall j, n, b$ are collected at the CU,
the CU can reconstruct all the sample covariance matrices according to \eqref{eq.sys.multi.cell} as follows
\begin{align}
\bar{\boldsymbol\Sigma}_b =\sum_{j=1}^B\mathbf{S}_j\bar{\mathbf{A}}_j^b\mathbf{G}_{bj}\mathbf{S}_j^H+\sigma_w^2\mathbf{I}, \,\forall b,
\end{align}
where $\bar{\mathbf{A}}_{j}^b=\operatorname{diag}\{\bar{a}_{j1}^b,\ldots,\bar{a}_{jN}^b\}$. Finally, the CU can perform cooperative activity detection based on the reconstructed sample covariance matrices $\bar{\boldsymbol\Sigma}_b, \forall b$ as follows
\begin{subequations}\label{eq.prob.multi.cu}
\begin{alignat}{2}\label{eq.prob.multi.cu.1}
&\underset{\{\mathbf{a}_{b}, \forall b\}}{\operatorname{minimize}}    &\quad& \sum_{b=1}^B \log\left|\boldsymbol\Sigma_b\right|+ \operatorname{tr}\left(\boldsymbol\Sigma_b^{-1}\bar{\boldsymbol\Sigma}_b\right)\\
&\operatorname{subject\,to} &      &a_{bn} \in [0,1], \,\forall b,n.
\label{eq.prob.multi.cu.2}
\end{alignat}
\end{subequations}

As compared to the problem \eqref{eq.prob2.multi} for infinite-capacity fronthaul links, the above problem formulation uses the reconstructed sample covariance matrices at the CU instead of the true sample covariance matrices. The rationale for employing this reconstruct-and-detect approach is that for a device, say device $n$ in cell $j$, it is difficult to directly aggregate the quantized activity indicator estimates $\bar{a}_{jn}^{b}, \forall b$ from the $B$ BSs since the accuracy levels of these estimates could be substantially different. Specifically, the results obtained from the BSs that are closer to the device are much more reliable, and the aggregation procedure should take the reliability into account, which is not straightforward. On the other hand, the quantized activity indicators preserve the essential information on device activity in the sample covariance matrix, so the reconstructed sample covariance matrix is a good approximation for device activity detection.

\section{Numerical Results}
\label{sec.simu}

In this section, we use simulation results to demonstrate the benefit of exploiting the knowledge of large-scale fadings. We consider a cellular network with hexagonal cells where BS-to-BS distance is $500$m and the potential users are uniformly distributed in the network. The channel path-loss is modeled as $128.1+37.6\log_{10}(d)$, where $d$ is BS-device distance in km. To ensure that the channel gains are bounded, we assume that the potential users are located at least $50$m away from the BS. The transmit power of each device is set as $23$dBm, and the background noise power is $-169$dBm/Hz over $10$MHz. We assume that all signature sequences have i.i.d.\ entries that are generated from a complex Gaussian distribution with zero mean and unit variance.

\subsection{Single-Cell Scenario}

\begin{figure}
    \centering
    \includegraphics[width=0.5\textwidth]{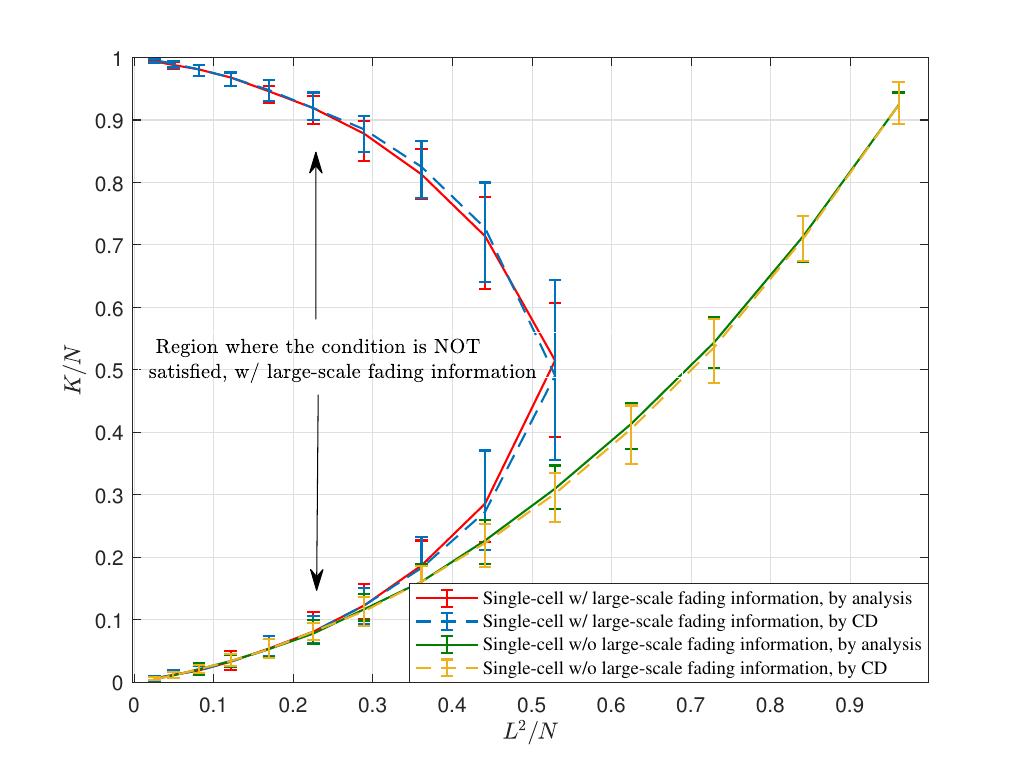}
  \caption{Phase transition of the covariance based activity detection in the single-cell scenario.}
  \label{fig.single.pt}
\end{figure}

We begin with the single-cell scenario with $N=1000$ devices uniformly distributed within the cell.
To obtain a numerical phase transition analysis of the covariance based activity detection, we numerically test the necessary and sufficient condition described in Theorem~\ref{single.condition.nolsf} or Theorem~\ref{single.condition.lsf} under a variety of choices of $L$ and $K$. In Fig.~\ref{fig.single.pt}, we draw the region of $(L^2/N, K/N)$ in which the condition is satisfied or not satisfied.
The result is obtained based on $100$ realizations of the random sequences and device activity patterns for each given $K$ and $L$.
The error bars indicate the range beyond which either all $100$ realizations or zero realization satisfy the condition. To validate the results of Theorem~\ref{single.condition.nolsf} and Theorem~\ref{single.condition.lsf}, we also use the coordinate descent (CD) algorithm to solve the detection problem in \eqref{eq.prob.single.nolsf} or \eqref{eq.prob.single.lsf} in the large $M$ limit, i.e., with ideal sample covariance matrix, to empirically plot a phase transition curve.

We observe from Fig.~\ref{fig.single.pt} that the curves obtained by the analysis and by the CD algorithm match very well. Moreover, Fig.~\ref{fig.single.pt} shows that the feasible region is much larger when the large-scale fading is exploited. This is mainly due to the symmetry of the phase transition curve with respect to $K/N=0.5$, which is implied by Proposition~\ref{theo3}.

\begin{figure}
    \centering
    \includegraphics[width=0.5\textwidth]{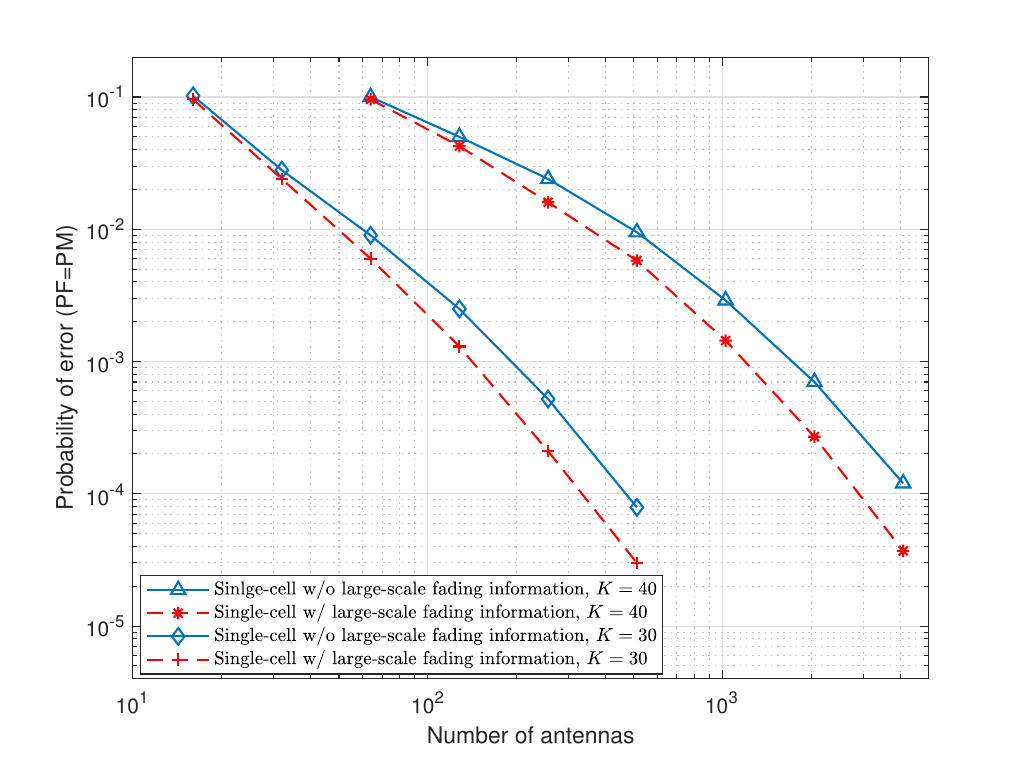}
  \caption{Performance of the activity detection in the single-cell scenario.}
  \label{fig.single.compare}
\end{figure}

The phase transition only provides the performance limit as the number of antennas goes to infinity. To compare performance under finite $M$, we plot the detection error against the number of antennas in Fig.~\ref{fig.single.compare}. We assume $L=25$ and $K=30, 40$ such that $(L^2/N=0.625, K/N=0.03, 0.04)$ lies in the feasible region for both cases with and without knowing the large-scale fadings. Note that there are two types of errors in activity detection: missed detection and false alarm, which can be traded off by adjusting the value of the threshold. For convenience, we select the threshold to achieve a point where two probabilities are equal. We observe from Fig.~\ref{fig.single.compare} that exploiting the large-scale fading information leads to a slightly better detection performance in this single-cell case.

\subsection{Multi-Cell Scenario with Infinite-Capacity Fronthauls}
\label{sec.simu.multi}

We consider a multi-cell system consisting of $7$ hexagonal cells with wrap-around at the boundary and $N=200$ devices uniformly distributed within each cell. In this subsection, the fronthauls between the BSs and the CU are assumed to have infinite capacity, so the sample covariance matrices are available at the CU exactly. 

\begin{figure}
    \centering
    \includegraphics[width=0.5\textwidth]{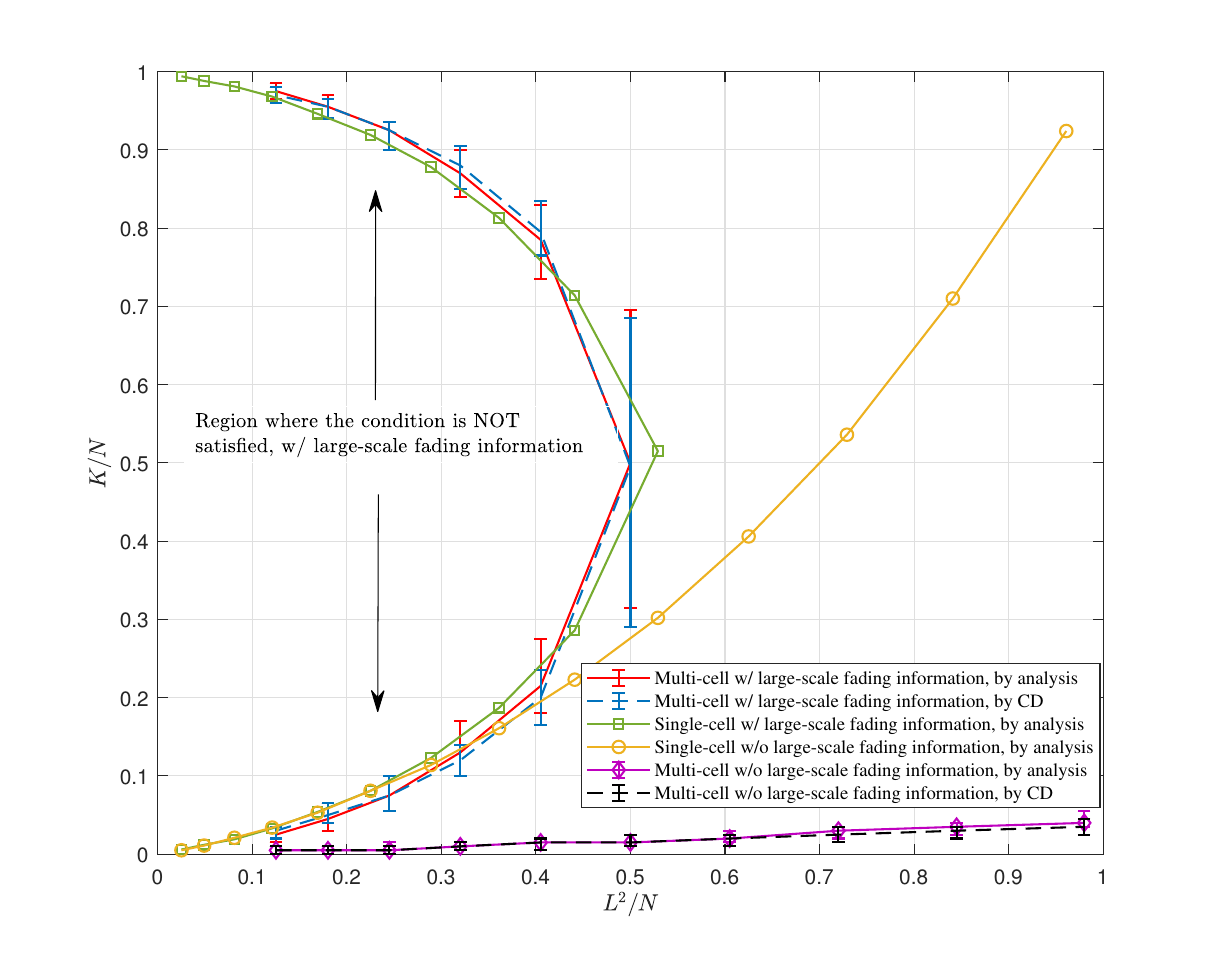}
  \caption{Phase transition of the covariance based activity detection in the multi-cell scenario with known large-scale fadings.}
  \label{fig.multi.pt}
\end{figure}

Fig.~\ref{fig.multi.pt} shows the numerical phase transition analysis of the covariance based activity detection in the multi-cell case. For comparison, we also plot the phase transition in the single-cell case. The multi-cell curves are obtained in a similar way as in Fig.~\ref{fig.single.pt}. For the case with known large-scale fadings, we numerically test the condition in Theorem~\ref{th.main.multi} under a variety of choices of $L$ and $K$ given $N=200$. We also use the CD algorithm to solve the detection problem in \eqref{eq.prob2.multi} with ideal sample covariance matrices to validate the phase transition curve. For the case with unknown large-scale fadings, the phase transition curve corresponds to the problem formulation in \eqref{eq.prob.multi.nolsf}. We observe from Fig.~\ref{fig.multi.pt} that the multi-cell curves obtained by the analysis and by the CD algorithm match well. We also observe that exploiting the large-scale fading information leads to much larger feasible region as compared to the case with unknown large-scale fadings in the multi-cell scenario. Moreover, with known large-scale fadings, the phase transition curves in the single-cell and the multi-cell cases are very similar, which means that the covariance based detection can work well in a similar region of $(L^2/N, K/N)$ in both single-cell and multi-cell cases as the number of antennas tends to infinity.

\begin{figure}
    \centering
    \includegraphics[width=0.5\textwidth]{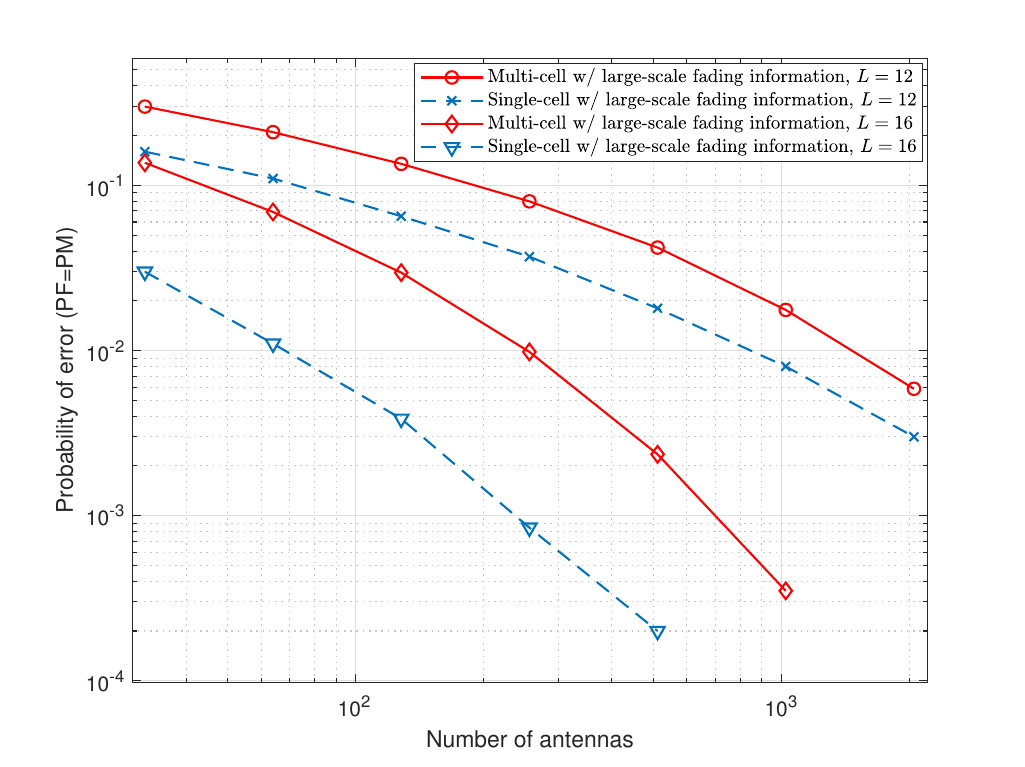}
  \caption{Performance comparison of single-cell and multi-cell cases with known large-scale fadings.}
  \label{fig.multi.compare.1}
\end{figure}

It should be mentioned that although the phase transition curves for single-cell and multi-cell cases are similar, as shown in Fig.~\ref{fig.multi.pt}, their detection performances may still differ when the number of antennas is finite. In Fig.~\ref{fig.multi.compare.1}, we compare the detection error probabilities in the multi-cell and single-cell cases with known large-scale fadings under finite $M$. We set $N=200$ and $K=20$ for both cases. We observe from Fig.~\ref{fig.multi.compare.1} that the covariance based detection method performs better in the single-cell case, and the performance gap becomes larger as $L$ increases.

\begin{figure}
    \centering
    \includegraphics[width=0.5\textwidth]{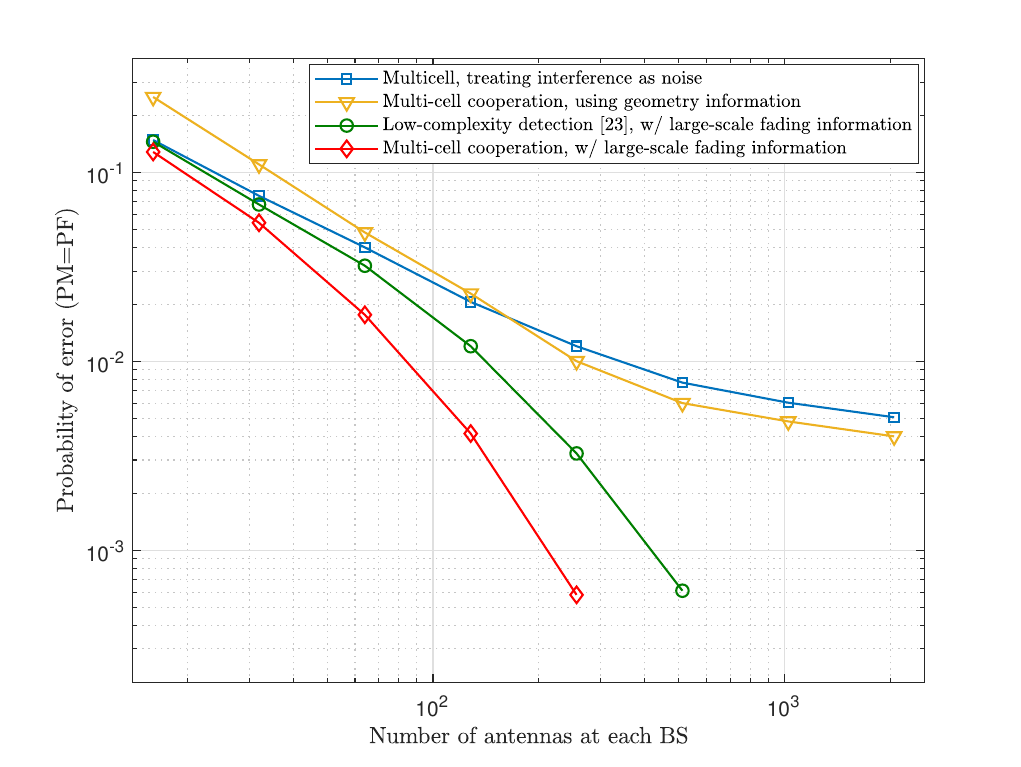}
  \caption{Performance comparison of different device activity detection strategies in the multi-cell case.}
  \label{fig.multi.compare.2}
\end{figure}

In Fig.~\ref{fig.multi.compare.2}, we consider three additional detection methods that can be used in the multi-cell case as benchmarks to illustrate the performance of the cooperative detection considered in this paper. These three methods are
\begin{itemize}
\item \emph{Treating Inter-cell Interference as Noise:} This is the scheme discussed in Section~\ref{sec.tin}, in which each BS operates independently to detect the active devices in the cell while treating inter-cell interference as noise. The covariance of the inter-cell interference is approximated by its time average.
\item \emph{Cooperative Detection Using Geometry of Location Information}: In this scheme, the CU determines the activity of device $n$ in cell $b$ by estimating $\gamma_{jbn}=a_{bn}g_{jbn}, j\in \mathcal{B}_{bn}$, where $\mathcal{B}_{bn}$ is the set of closest BSs to this device. Note that the values of $\gamma_{jbn}$ across $j\in \mathcal{B}_{bn}$ have interdependencies because of the device and the BS locations. To perform cooperative detection, we assume that the CU has a radio map of all possible values of $\gamma_{jbn}$, $j\in \mathcal{B}_{bn}$, based on the possible device locations. We then use CD to estimate $\gamma_{jbn}, j\in \mathcal{B}_{bn}$. Specifically, we update $\gamma_{jbn}, j\in \mathcal{B}_{bn}$ at each step simultaneously by using exhaustive searching on this radio map (or possibly declaring that the user is not active). We use $|\mathcal{B}_{bn}|=3$, i.e., three closest BSs, in the simulations.
\item \emph{Low Complexity Detection with Known Large-Scale Fadings \cite{Ganesan2020}}: The CU estimates $a_{bn}$ iteratively based on the sample covariance matrix at a single BS $i$ that has the largest channel gain, i.e., $i=\arg\max_{j}\{g_{jbn}\}$.
\end{itemize}

We assume $N=200$, $K=20$ and $L=20$. We observe from Fig.~\ref{fig.multi.compare.2} that the cooperative activity detection scheme proposed in this paper outperforms the above three strategies. We also observe that without knowing the large-scale fadings, the improvement by the cooperative detection scheme based on the geometry of location information is marginal as compared to simply treating inter-cell interference as noise.

\subsection{Multi-Cell Scenario with Capacity-Limited Fronthauls}

Finally, we study the impact of quantization for capacity-limited fronthauls. We compare two quantization schemes: i) quantization of the sample covariance matrix; and ii) quantization of the activity indicators after a preliminary detection at each BS, as in Section~\ref{sec.quan}. We consider the multi-cell setup with $N=200$, $K=20$, and $L=20$.

\begin{figure}
    \centering
    \includegraphics[width=0.5\textwidth]{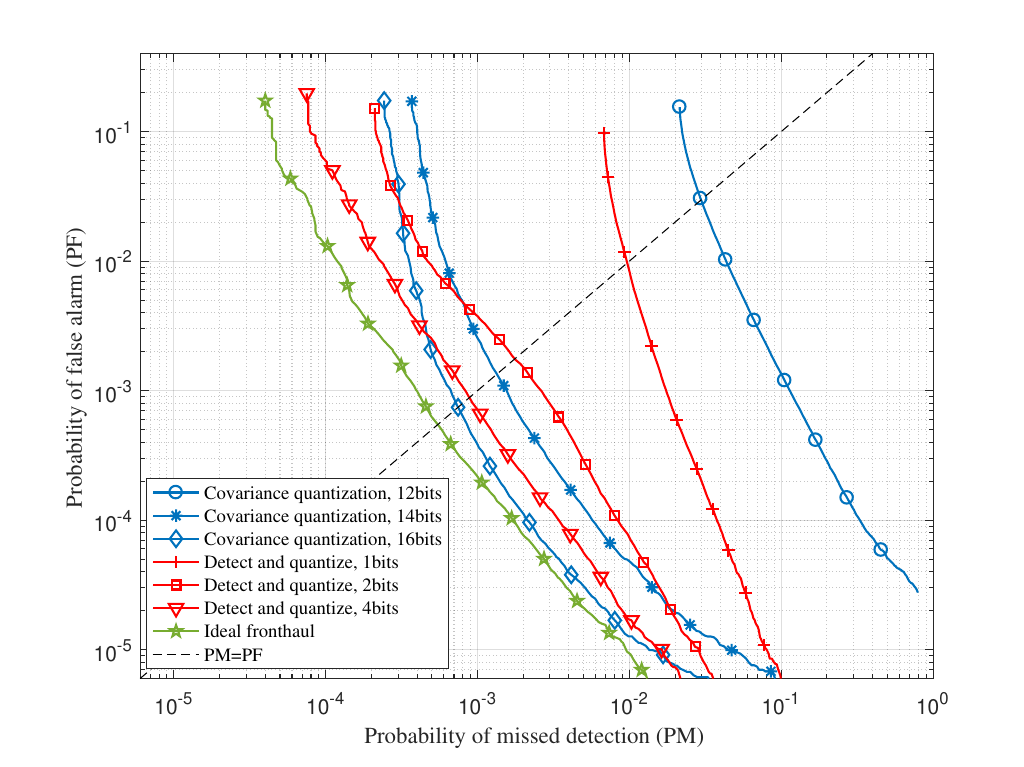}
  \caption{Performance comparison of two quantization schemes for multi-cell activity detection.}
  \label{fig.multi.q.1}
\end{figure}

Fig.~\ref{fig.multi.q.1} shows how many quantization bits per real-valued scalar are needed to approach the infinite-capacity fronthaul result.
We observe from Fig.~\ref{fig.multi.q.1} that for the sample covariance matrix quantization, at least $14$ bits are needed to achieve a performance close to the infinite-capacity fronthaul case, whereas for the activity indicator quantization, a similar performance can be achieved with $2$ bits per scalar. 

\begin{figure}
    \centering
    \includegraphics[width=0.5\textwidth]{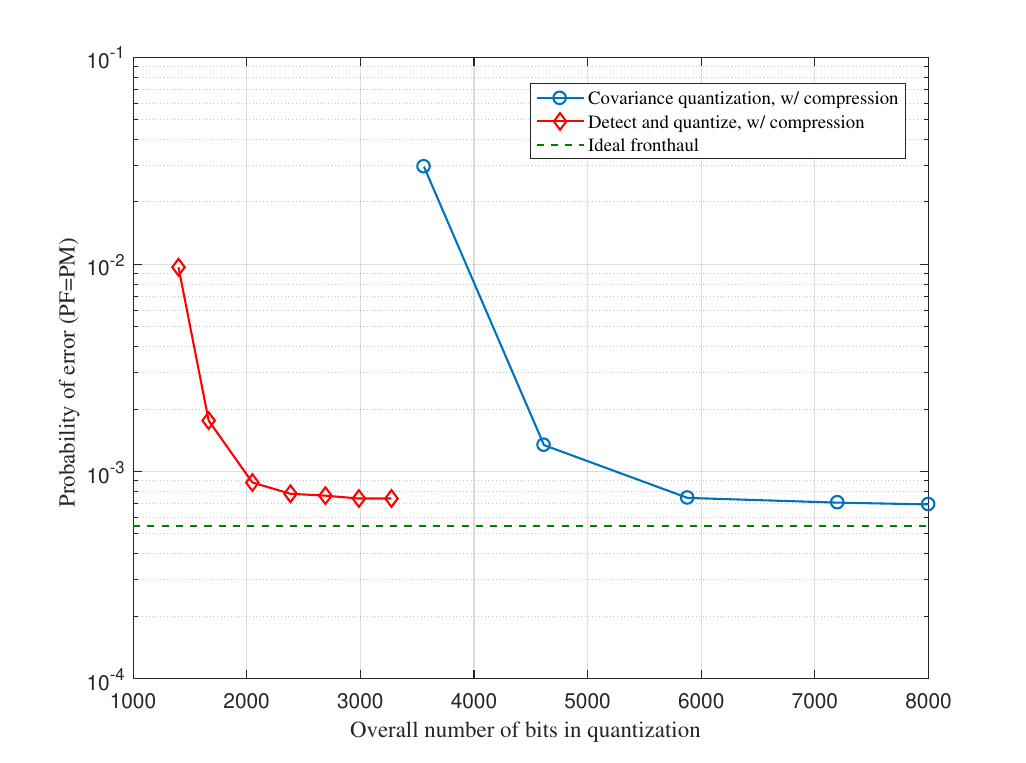}
  \caption{Comparison of overall number of bits needed in these two quantization schemes.}
  \label{fig.multi.q.2}
\end{figure}

Note that as there are more scalars to be quantized in the activity indicator quantization scheme, for fair comparison, we should sum up the total number of quantization bits.
Further, the quantization bits can be compressed via entropy coding.
For the activity indicator quantization scheme, due to the fact that most of the activity indicators are zero or one, they are highly compressible. In contrast, the quantization bits in the sample covariance matrix quantization scheme cannot be compressed much further. In Fig.~\ref{fig.multi.q.2}, we compare the overall number of quantization bits needed to approach the case with infinite-capacity fronthauls. We use Huffman coding for compression in both cases.
We observe that 
quantizing the activity indicators can substantially reduce the fronthaul requirements as compared to quantizing the sample covariance matrices.

\section{Conclusion}
\label{sec.conclusion}

This paper studies the benefit of exploiting the large-scale fading information
for device activity detection in a cooperative multi-cell massive MIMO system.
We use a covariance based activity detection method and present a
numerical phase transition analysis that quantifies the required length of
the signature sequences to ensure reliable detection for a given number of
(active) devices in the massive MIMO regime.
The analysis shows that the required sequence length in the multi-cell case can be
made to be almost the same as that in the single-cell case by cooperative detection in a C-RAN architecture, but only if the large-scale fading information is exploited.
This paper also proposes a novel quantization design over the fronthaul of the C-RAN that includes a preliminary detection at the BSs and a covariance reconstruction at the CU for
the case where the fronthauls have finite-capacity constraints.
As compared to directly quantizing the sample covariance
matrices, the proposed scheme can significantly save fronthaul resources.

\bibliographystyle{IEEEtran}
\bibliography{IEEEabrv,chenbib}

\begin{thebibliography}{10}
\providecommand{\url}[1]{#1}
\csname url@samestyle\endcsname
\providecommand{\newblock}{\relax}
\providecommand{\bibinfo}[2]{#2}
\providecommand{\BIBentrySTDinterwordspacing}{\spaceskip=0pt\relax}
\providecommand{\BIBentryALTinterwordstretchfactor}{4}
\providecommand{\BIBentryALTinterwordspacing}{\spaceskip=\fontdimen2\font plus
\BIBentryALTinterwordstretchfactor\fontdimen3\font minus
  \fontdimen4\font\relax}
\providecommand{\BIBforeignlanguage}[2]{{%
\expandafter\ifx\csname l@#1\endcsname\relax
\typeout{** WARNING: IEEEtran.bst: No hyphenation pattern has been}%
\typeout{** loaded for the language `#1'. Using the pattern for}%
\typeout{** the default language instead.}%
\else
\language=\csname l@#1\endcsname
\fi
#2}}
\providecommand{\BIBdecl}{\relax}
\BIBdecl

\bibitem{Bockelmann2016}
C.~Bockelmann, N.~Pratas, H.~Nikopour, K.~Au, T.~Svensson, C.~Stefanovic,
  P.~Popovski, and A.~Dekorsy, ``Massive machine-type communications in 5{G}:
  Physical and {MAC}-layer solutions,'' \emph{IEEE Commun. Mag.}, vol.~54,
  no.~9, pp. 59--65, Sept. 2016.

\bibitem{Dawy2017}
Z.~{Dawy}, W.~{Saad}, A.~{Ghosh}, J.~G. {Andrews}, and E.~{Yaacoub}, ``Toward
  massive machine type cellular communications,'' \emph{IEEE Wireless Commun.},
  vol.~24, no.~1, pp. 120--128, Feb. 2017.

\bibitem{Chen2021JSAC}
X.~Chen, D.~W.~K. Ng, W.~Yu, E.~G. Larsson, N.~Al-Dhahir, and R.~Schober,
  ``Massive access for 5{G} and beyond,'' \emph{IEEE J.\ Sel.\ Areas Commun.},
  vol.~39, no.~3, pp. 615--637, Mar. 2021.

\bibitem{Liu2018b}
L.~Liu, E.~G. Larsson, W.~Yu, P.~Popovski, {\v C}.~Stefanovi{\'c}, and
  E.~de~Carvalho, ``Sparse signal processing for grant-free massive
  connectivity: A future paradigm for random access protocols in the {I}nternet
  of {T}hings,'' \emph{IEEE Signal Process. Mag.}, vol.~35, no.~5, pp. 88--99,
  Sept. 2018.

\bibitem{Liu2018}
L.~Liu and W.~Yu, ``Massive connectivity with massive {MIMO} ---{P}art {I}:
  Device activity detection and channel estimation,'' \emph{IEEE Trans. Signal
  Process.}, vol.~66, no.~11, pp. 2933--2946, June 2018.

\bibitem{Chen2018}
Z.~Chen, F.~Sohrabi, and W.~Yu, ``Sparse activity detection for massive
  connectivity,'' \emph{IEEE Trans. Signal Process.}, vol.~66, no.~7, pp.
  1890--1904, Apr. 2018.

\bibitem{Haghighatshoar2018}
S.~Haghighatshoar, P.~Jung, and G.~Caire, ``Improved scaling law for activity
  detection in massive {MIMO} systems,'' in \emph{Proc. IEEE Int. Symp. Inf.
  Theory (ISIT)}, Vail, CO, USA, June 2018, pp. 381--385.

\bibitem{Fengler2019a}
A.~Fengler, S.~Haghighatshoar, P.~Jung, and G.~Caire, ``Non-{B}ayesian activity
  detection, large-scale fading coefficient estimation, and unsourced random
  access with a massive {MIMO} receiver,'' \emph{IEEE Trans.\ Inf.\ Theory},
  vol.~67, no.~5, pp. 2925--2951, May 2021.

\bibitem{ChenZ2020}
\BIBentryALTinterwordspacing
Z.~Chen, F.~Sohrabi, Y.-F. Liu, and W.~Yu, ``Phase transition analysis for
  covariance based massive random access with massive {MIMO},'' 2020. [Online].
  Available: \url{https://arxiv.org/abs/2003.04175}
\BIBentrySTDinterwordspacing

\bibitem{Donoho2009}
D.~Donoho, A.~Maleki, and A.~Montanari, ``Message-passing algorithms for
  compressed sensing,'' \emph{Proc. Nat. Acad. Sci.}, vol. 106, no.~45, pp.
  18\,914--18\,919, Nov. 2009.

\bibitem{Sun2019}
Z.~{Sun}, Z.~{Wei}, L.~{Yang}, J.~{Yuan}, X.~{Cheng}, and L.~{Wan},
  ``Exploiting transmission control for joint user identification and channel
  estimation in massive connectivity,'' \emph{IEEE Trans. Commun.}, vol.~67,
  no.~9, pp. 6311--6326, Sept. 2019.

\bibitem{Senel2018}
K.~Senel and E.~G. Larsson, ``Grant-free massive {MTC}-enabled massive {MIMO}:
  A compressive sensing approach,'' \emph{IEEE Trans. Commun.}, vol.~66,
  no.~12, pp. 6164--6175, Dec. 2018.

\bibitem{Jiang2020}
S.~Jiang, X.~Yuan, X.~Wang, C.~Xu, and W.~Yu, ``Joint user identification,
  channel estimation, and signal detection for grant-free {NOMA},'' \emph{IEEE
  Trans.\ Wireless Commun.}, vol.~19, no.~10, pp. 6960--6976, Oct. 2020.

\bibitem{Simeone2016}
Z.~Utkovski, O.~Simeone, T.~Dimitrova, and P.~Popovski, ``Random access in
  {C-RAN} for user activity detection with limited-capacity fronthaul,''
  \emph{IEEE Signal Process. Lett.}, vol.~24, no.~1, pp. 17--21, Jan. 2017.

\bibitem{Chen2019c}
Z.~{Chen}, F.~{Sohrabi}, and W.~{Yu}, ``Multi-cell sparse activity detection
  for massive random access: Massive {MIMO} versus cooperative {MIMO},''
  \emph{IEEE Trans. Wireless Commun.}, vol.~18, no.~8, pp. 4060--4074, Aug.
  2019.

\bibitem{Ke2020}
M.~Ke, Z.~Gao, Y.~Wu, X.~Gao, and K.-K. Wong, ``Massive access in cell-free
  massive {MIMO}-based {I}nternet of {T}hings: Cloud computing and edge
  computing paradigms,'' \emph{IEEE J. Sel. Areas Commun.}, vol.~39, no.~3, pp.
  756--772, Mar. 2021.

\bibitem{Lau2015}
X.~Xu, X.~Rao, and V.~K.~N. Lau, ``Active user detection and channel estimation
  in uplink {CRAN} systems,'' in \emph{Proc. IEEE Int. Conf. Commun. (ICC)},
  London, UK, June 2015, pp. 2727--2732.

\bibitem{Ahn2019}
J.~{Ahn}, B.~{Shim}, and K.~B. {Lee}, ``{EP}-based joint active user detection
  and channel estimation for massive machine-type communications,'' \emph{IEEE
  Trans. Commun.}, vol.~67, no.~7, pp. 5178--5189, July 2019.

\bibitem{Shao2020}
X.~{Shao}, X.~{Chen}, and R.~{Jia}, ``A dimension reduction-based joint
  activity detection and channel estimation algorithm for massive access,''
  \emph{IEEE Trans. Signal Process.}, vol.~68, no.~1, pp. 420--435, Jan. 2020.

\bibitem{Chen2019a}
Z.~Chen, F.~Sohrabi, Y.-F. Liu, and W.~Yu, ``Covariance based joint activity
  and data detection for massive random access with massive {MIMO},'' in
  \emph{Proc. IEEE Int. Conf. Commun. (ICC)}, Shanghai, China, May 2019, pp.
  1--6.

\bibitem{Dong2020}
\BIBentryALTinterwordspacing
J.~Dong, J.~Zhang, Y.~Shi, and J.~H. Wang, ``Faster activity and data detection
  in massive random access: A multi-armed bandit approach,'' 2020. [Online].
  Available: \url{https://arxiv.org/abs/2001.10237}
\BIBentrySTDinterwordspacing

\bibitem{ShaoUnsourced2020}
X.~Shao, X.~Chen, D.~W.~K. Ng, C.~Zhong, and Z.~Zhang, ``Cooperative activity
  detection: Sourced and unsourced massive random access paradigms,''
  \emph{IEEE Trans.\ Signal Process.}, vol.~68, pp. 6578--6593, Nov. 2020.

\bibitem{Ganesan2020}
U.~K. Ganesan, E.~Bj{\"o}rnson, and E.~G. Larsson, ``An algorithm for
  grant-free random access in cell-free massive {MIMO},'' in \emph{Proc. IEEE
  Workshop Signal Process. Adv. Wireless Commun. (SPAWC)}, Atlanta, GA, USA,
  May 2020, pp. 1--5.

\bibitem{Jiang2020a}
D.~Jiang and Y.~Cui, ``{ML} estimation and {MAP} estimation for device
  activities in grant-free random access with interference,'' in \emph{Proc.\
  IEEE Wireless Commun.\ Netw.\ Conf.\ (WCNC)}, Seoul, South Korea, May 2020,
  pp. 1--6.

\end{thebibliography}

\end{document}